\documentclass[twocolumn,prd,showpacs,preprintnumbers,nofootinbib]{revtex4-1}
\usepackage{amsmath,amssymb,pgf,graphicx,color,url}
\newcommand{\be}{\begin{equation}}
\newcommand{\ee}{\end{equation}}
\newcommand{\g}{\gamma}

\usepackage[unicode=true,pdfusetitle,
bookmarks=true,bookmarksnumbered=true,bookmarksopen=true,bookmarksopenlevel=1,
breaklinks=false,pdfborder={0 0 0},backref=false,colorlinks=true]
{hyperref}
\hypersetup{
	citecolor=blue,filecolor=blue,linkcolor=blue,urlcolor=blue}

\begin{document}

\preprint{INR-TH-2016-016}

\title{Constraining the production of cosmic rays by pulsars}

\author{
Mikhail\,M.\,Ivanov$^{1,2,3}$,
Maxim\,S.\,Pshirkov$^{4,1,5}$,
and Grigory\,I.\,Rubtsov$^{1}$} 
\affiliation{ 
$^1$ Institute  for Nuclear Research of the Russian Academy of Sciences, \normalsize\it 117312 Moscow, Russia\\
$^2$ Faculty of Physics, Moscow State University, \normalsize\it 119991 Moscow, Russia \\
$^3$ FSB/IPHYS/LPPC, \'Ecole Polytechnique F\'ed\'erale de Lausanne, \normalsize\it CH-1015 Lausanne, Switzerland\\
$^4$ Sternberg Astronomical Institute, Lomonosov Moscow State University, \normalsize\it 119992, Moscow, Russia\\
$^5$ Pushchino Radio Astronomy Observatory, \normalsize\it 142290 Pushchino, Russia
}

\begin{abstract}
One of the possible sources of hadronic cosmic rays (CRs) are newborn pulsars.
If this is indeed the case, 
they should feature diffusive gamma-ray halos produced by interactions of CRs with interstellar gas.
In this paper we try to identify extended gamma-ray emission around young pulsars,
making use of the 7-year Fermi-LAT data.
For this purpose we select and analyze a set of eight pulsars 
that are most likely to possess detectable gamma-ray halos.
We find extended emission that might be interpreted as a gamma-ray halo
only in the case of PSR J0007+7303.
Its luminosity accords with the total energy of injected cosmic rays $\sim 10^{50}$ erg,
although other interpretations of this source are possible.
Irrespectively of the nature of this source we put
bounds on the luminosity of gamma-ray halos which
suggest that pulsars' contribution
to the overall energy budget of galactic CRs is subdominant 
in the GeV-TeV range.
\end{abstract}
\maketitle

\section{Introduction}

Cosmic ray (CR) experiments have 
allowed for the measurement of the
spectrum and chemical composition of galactic CRs. 
The observed value of the latter requires an average cumulative  
power of CR sources $L_{CR}\sim 10^{41}$ erg/s \cite{Strong:2010pr} at energies 
$E_{CR}>0.1$~GeV.

The bulk of galactic cosmic rays is widely believed to originate from 
supernova remnants (SNRs), 
see the recent reviews \cite{Blasi:2013rva,Amato:2014xua}. 
This hypothesis is supported by a number of 
convincing, independent and yet circumstantial indications.
The most remarkable are
recent observations of the SNRs W44 and IC433 \cite{Tavani:2010vf,Ackermann:2013wqa,Cardillo:2014kaa} 
which allowed for
confident conclusions on the hadronic nature of their gamma-ray emission.
Nevertheless, there are still puzzles to be resolved, 
e.g., a mismatch between the predicted and 
observed slopes of the gamma-ray spectrum. 
It is also not clear whether SNRs indeed accelerate CRs up to the ``knee"
energies $\sim 10^{6}$ GeV. 
Besides, several breaks observed in the Galactic CR spectrum 
\cite{Neronov:2011wi,Adriani:2011cu}
hint at the existence of multiple components in the interstellar CR flux.
This motivates a search for some complementary scenarios of CR production.
CRs can be produced by mechanisms operating at large scales,
such as acceleration in superbubbles \cite{Ackermann:2011lfa,Bykov:2001}
or Galactic-wind shocks \cite{Butt:2009zz}. 
This scenario is supported by 
the chemical composition of the low-energy cosmic ray flux 
\cite{Wiedebeck:1999} and by the extended gamma-ray emission 
observed in the Cygnus superbubble~\cite{Ackermann:2011lfa}. 

Alternatively, pulsars and their pulsar wind nebulae (PWNe) could be viable sources of CRs
\cite{Gunn:1969me,Fang:2012rx,Lemoine:2014ala,Amato:2003kw}.
Indeed, the rotation energy of neutron stars at birth 
is sufficient to produce the required CR power \cite{Ostriker:1969me,Neronov:2012kz}.
It is well established that the rotation energy of young pulsars 
is spent extremely efficiently on the production and acceleration of leptons 
\cite{Hooper:2008kg,Bednarek:2003tk,Amato:2013fua}.
Furthermore, the most successful theoretical models of particle 
acceleration at pulsar winds \cite{Hoshino:1992zz,Gallant:1994zz}
predict that ions should typically carry energy larger than that of electrons and positrons. 
However, the emission associated with high-energy leptons
may introduce a serious obstacle to testing the production of CRs by pulsars:
the hadron component of the gamma-ray flux could be deeply hidden 
in the overwhelming emission of leptonic origin.

Fortunately, there is a potential way out of this predicament.
When CRs escape their
sources they should interact with interstellar gas and produce 
observable gamma-ray emission.
According to an estimate given below 
a typical size of an extended 
halo around a young pulsar should be $\sim 100$ pc.
Unlike ions, leptons undergo severe energy losses due to synchrotron emission and inverse Compton
scattering. Thus, one may expect that at distances comparable to the halo size
the energy density of leptons becomes suppressed \cite{Aharonian:1995zz},
and gamma-ray emission is dominated by the hadronic component.

Several  candidates for the extended gamma-ray halos 
around young pulsars were found in Ref. \cite{Neronov:2012kz},
which may be considered as an evidence in favor of CR production by pulsars.
Moreover, the results obtained in Ref. \cite{Neronov:2012kz} led to 
the conclusion that gamma-ray halos should exist around nearly all young pulsars
with a spin-down age $T_{SD}\lesssim$ 30 kyr.
The observations of the very high-energy neutrinos reported by IceCube \cite{Aartsen:2013bka} 
can also be consistently interpreted within this scenario
\cite{Neronov:2013lza,Tchernin:2013wfa}.
All of these pieces of evidence and 
their relevance for unveiling the puzzles of CRs suggest that
the hypothesis of CR production by pulsars
requires further investigation, which we perform in this paper.

This paper is organized as follows. 
In Sec. \ref{sec:theory} we discuss theoretical aspects of CR production by pulsars 
and properties of hypothetical gamma-ray halos around them.
Section \ref{sec:sample} is devoted to the selection of pulsars for further tests.
In Sec. \ref{sec:method} we discuss the analysis of the Fermi-LAT data. 
In Sec. \ref{sec:simul} we discuss the dependence of the statistical significance 
on halo fluxes retrieved from simulations. This will allow us to constrain the halo luminosity.
Section \ref{sec:results} is devoted to the analysis of the 
selected pulsars with the Fermi-LAT  data. 
The results are interpreted in Sec. \ref{sec:disc}.
We draw conclusions in Sec. \ref{sec:concl}.
In Appendix \ref{app:NSpuls} we describe properties 
of the pulsars from Ref.~\cite{Neronov:2012kz}, whereas Appendix \ref{app:simul}
contains the details of the simulations. In Appendix \ref{app:fluct}
we verify that the sources found in Ref. \cite{Neronov:2012kz} are not the result of 
statistical fluctuations. 
Finally, in Appendix \ref{app:3fgl} we show best fits for the sources
from our analysis and compare them to the values from the 3FGL catalogue.

\section{Theoretical preliminaries}
\label{sec:theory}

In order to reproduce the observed density 
of CRs at Earth $\sim 1$ eV/cm$^{3}$ \cite{Neronov:2011wi}
one requires the following total time-averaged luminosity of CR sources \cite{Strong:2010pr},
\be
\label{eq:crpower}
 L_{CR}^{tot}\simeq 8\times 10^{40}\; \frac{\mathrm{erg}}{\mathrm{s}}\,.
\ee
Notice that this is the total power of hadrons and nuclei with kinetic
energies $E_{CR}\gtrsim 0.1$ GeV.

The most plausible sources of this power are supernovae explosions, 
which release $\sim 10^{51}$ erg with the rate ($1/30-1/130$) yr$^{-1}$ \cite{Diehl:2006cf}. 
Indeed, a rough estimate implies that some $\sim 10\%$ of this energy
would totally account for the bulk of galactic cosmic rays,
\be
\label{eq:SNrate}
\begin{split}
&L_{SN}^{CR}\sim \mathcal{E}^{tot}_{CR} \mathcal{R}_{SN}\\
&\simeq 10^{41}\; \frac{\mathrm{erg}}{\mathrm{s}}
\left[\frac{\mathcal{E}^{tot}_{CR}}{2\times 10^{50}\;\mathrm{erg}}\right]\left[\frac{\mathcal{R}_{SN}}{1/50\;\mathrm{ yr}^{-1}}\right]\,,
\end{split}
\ee
where $\mathcal{E}^{tot}_{CR}$ is the total energy output per supernova in the form of CRs.

As another option, the required energy input can be provided by
fast-spinning newborn pulsars \cite{Ostriker:1969me,Neronov:2012kz},
which possess a sufficient amount of rotational energy,
\be
\label{eq:erot}
\begin{split}
& E_{rot}= \frac{I_{NS}\Omega^2}{2}  \\
& \simeq 2\times 10^{50} \; \mathrm{erg}\left[\frac{I_{NS}}{10^{45}\; \mathrm{g cm}^2}\right]
\left[\frac{10\; \mathrm{ms}}{P_{ini}}\right]^2\,.
\end{split}
\ee
In fact, theoretical models predict 
that initial periods of neutron stars at birth can be even
shorter than 1 ms in the absence of strong magnetic coupling  
between a stellar core and outer layers \cite{Ott:2005wh}.

The pulsar birthrate  
should typically be smaller than the core-collapsed supernova rate; 
thus, if young pulsars are the only source of CRs
they should inject more CRs than is expected from supernovae. 
Unfortunately, current measurements of the pulsar birthrate are less certain than those of the supernova rate
\cite{Vranesevic:2003tp,Keane:2008jj}. Hence, we stick to the latter in this paper. 
The uncertainly in the supernova rate induces a significant scatter over the required energy,
\be
\label{eq:pulsarate}
\mathcal{E}^{tot}_{CR}\simeq (1-5)\times 10^{50} \text{erg}\,.
\ee

When released, the cosmic rays generated by a pulsar 
diffuse away through the Galactic magnetic field and
fill a spherical volume whose radius can be estimated as
\be
\label{eq:diff}
r_{CR}\simeq  2\sqrt{D T_{SD}}\,,
\ee
where $D$ is the energy-dependent diffusion coefficient and 
$T_{SD}$ is the pulsar's spin-down age, which can be taken as an estimate 
for the typical time passed since the CRs' emission. 
The diffusion coefficient $D$ is given by
\cite{Strong:2010pr,Blasi:2011fi}
\be
\label{eq:diffcoeff}
D=D_{28}\times  10^{28} \left[\frac{E_{CR}}{3 \;\mathrm{GeV}}\right]^{\delta} \mathrm{cm}^2\mathrm{/s}\,, \quad \delta=0.4\pm 0.1\,,
\ee
where the prefactor $D_{28}\sim 1$ and
we assumed that the rigidity of CRs is the same as that of protons.
Notice that the uncertainty of the prefactor $D_{28}$ 
is up to a factor of~$3$.
The size of the CR halo around a pulsar \eqref{eq:diff} is given by\footnote{In Ref. \cite{Neronov:2012kz} the same estimate yielded a slightly smaller distance $r_s= 80$ pc. 
However, this numerical inaccuracy does not alter any results.}
\be
\label{eq:size2}
r_{CR}\simeq  120\times D^{1/2}_{28} 
\left[\frac{T_{SD}}{10\;\mathrm{kyr}}\right]^{1/2}
\left[\frac{E_{CR}}{1\;\mathrm{TeV}}\right]^{0.2}\;\mathrm{pc}\,.
\ee

As CRs interact with the interstellar 
medium, the CR halo should have a gamma-ray counterpart. 
In what follows we will 
assume a typical energy yield in gamma rays 
$\kappa~\approx~0.2$. This value for the yield is shown to agree quite well 
with precise numerical calculations \cite{Aharonian:2000iz,Kelner:2006tc}. 

One finds that the characteristic angular size of the gamma-ray
halo scales with $T_{SD}$, the photon energy $E_{\gamma}=\kappa E_{CR}$, 
and the distance to the source $r_s$ as
\be
\label{eq:size1}
\begin{split}
R_{halo}&=\frac{r_{CR}}{r_s} \\
& \simeq 1.4^\circ D^{1/2}_{28} \left[\frac{5\;\mathrm{kpc}}{r_s}\right]\left[\frac{T_{SD}}{10\;\mathrm{kyr}}\right]^{1/2}\left[\frac{E_{\gamma}}{200\;\mathrm{GeV}}\right]^{0.2}\,.
\end{split}
\ee
In what follows we will use 
uppercase letters $R$ to denote angular
distances and lowercase letters $r$ to denote physical distances.

The protons and nuclei produce gamma rays in inelastic
collisions with interstellar nucleons mostly due to the production
and subsequent decay of $\pi^0$ mesons.
The cross-section for the inelastic {\it pp} scattering 
has a logarithmic dependence on $E_{CR}$ and
declines abruptly
at energies $E_{CR}\lesssim 2$ GeV \cite{PDG}.
Thus, the resulting gamma-ray spectrum of a halo should be dominated by photons 
with energies $E_\g \gtrsim 0.5$ GeV.
Nevertheless, we will see in what follows
that the gamma-ray halos can be unambiguously 
detected only at energies $E_\gamma\gtrsim 1$ GeV. 
Hence, the relevant energy range of CRs contributing to this emission is $E_{CR}\gtrsim$ 5 GeV.

The luminosity of this halo can be
estimated using a typical interaction time of CRs in the interstellar medium (ISM),
\be
\label{eq:tint}
t_{int}=\frac{1}{c\sigma_{pp}n_{ISM}} \simeq  3\times 10^7 \left[\frac{1\; \mathrm{cm}^{-3}}{n_{ISM}}\right]\;\mathrm{yr},
\ee
where we have taken $\sigma_{pp}= 3\times 10^{-26}$ cm$^2$ 
as an average cross section
for the inelastic {\it pp} scattering 
for protons with $E_{CR}>$ 5 GeV 
\cite{PDG}, and the {\it average} interstellar matter density in the Galactic disc is $n_{ISM}\sim 1\;$cm$^{-3}$
\cite{Korchagin:2003yk}.
Making use of Eq.~\eqref{eq:tint}, one obtains the following halo luminosity:
\be
\label{eq:halolum}
\begin{split}
& L_{\g}^{E_\g \gtrsim 1\;\mathrm{GeV}} \sim \kappa\frac{\mathcal{E}^{halo}_{CR}}{t_{int}} \\
& \simeq 4\times 10^{34}\left[\frac{\kappa}{0.2}\right]
\left[\frac{\mathcal{E}^{halo}_{CR}}{2\times 10^{50}\;\mathrm{erg}}\right] \left[\frac{n_{ISM}}{1\; \mathrm{cm}^{-3}}\right]
\frac{\mathrm{erg}}{\mathrm{s}}\,,
\end{split}
\ee
where by $\mathcal{E}^{halo}_{CR}$
we denoted the total energy of cosmic rays with $E_{CR}\gtrsim 5$ GeV,
injected by a pulsar.

It should be pointed out that accurate numerical calculations \cite{Aharonian:2000iz,Kelner:2006tc,Strong:2004de}
imply that for realistic CR spectra  
the spectrum of produced gamma rays has a maximum at $E_\g \simeq 1$ GeV
and drops sharply at lower energies. 
Thus, 
one can think of $\mathcal{E}^{halo}_{CR}$ 
as the total energy of {\it all} CRs produced by a pulsar.

The candidates for the gamma-ray halos around pulsars
were found in Ref. \cite{Neronov:2012kz}
using the 3-year Fermi-LAT data above 100 GeV. 
These candidates will be referred to as {\it N-S} sources in what follows.
The {\it N-S} sources are listed in Table II of Ref. \cite{Neronov:2012kz}
and have the following typical fluxes:
\be
\label{eq:flux}
F^{E_\g>100~ \mathrm{GeV}}\simeq 5\times 10^{-11}\left[\frac{5 \;\mathrm{kpc}}{r_s}\right]^2 \frac{\mathrm{erg}}{\mathrm{cm}^2 \cdot \mathrm{s}} \,,
\ee
Assuming a power-law spectrum
of photons with $\Gamma=2$, one obtains\footnote{
We adopt $\Gamma=2$ here 
in order to obtain a conservative 
estimate for the total luminosity.} 
the following flux above $1$ GeV:
\be
F^{E_{\gamma}\geq 1\;\mathrm{GeV}}\simeq 2\times 10^{-10}\left[\frac{5 \;\mathrm{kpc}}{r_s}\right]^2 \frac{\mathrm{erg}}{\mathrm{cm}^2\cdot \mathrm{s}}\,,
\ee
yielding the luminosity
\be 
\label{eq:nslum}
L_{\g}^{E_\g\geq 1\;\mathrm{GeV}}=F^{E_{\gamma}\geq 1\;\mathrm{GeV}}4\pi r_s^2
\simeq 6\times  10^{35} \;\frac{\mathrm{erg}}{ \mathrm{s}}\,,
\ee
which is 20 times larger than our estimate \eqref{eq:halolum}.
This mismatch can explained by the fact that almost all of the {\it N-S} sources are
situated in the Norma arm, a peculiar star-forming region with a high-density interstellar medium.
This point will be discussed in more detail in Sec.\ref{sec:disc}.

There can be several difficulties 
with the identification of gamma-ray halos in data, e.g.,
an overlap with other gamma-ray sources and
background uncertainties.
Postponing for a moment statistical and instrumental ambiguities (to be discussed later),
we focus now on some theoretical issues 
which can have an impact on observations.

Pulsars are often located in the vicinity of SNR shells,
many of which are associated with extended gamma-ray sources.
Thus, one might worry about the disentanglement between 
SNRs and gamma-ray halos.
The SNRs, however, have much smaller 
angular extension compared to CR halos. Indeed, in the adiabatic 
Sedov-Taylor phase \cite{Taylor,Sedov} 
the SNR radius can be estimated as
\be
r_{SNR}\approx \left(\frac{25 \mathcal{E}_{SN}}{4\pi \rho_0}\right)^{0.2}t^{0.4}\,,
\ee
where $\mathcal{E}_{SN}$ is the energy of the supernova explosion and $\rho_0$
is the preexplosion density of the interstellar medium. This implies that the observed 
angular size of 
the supernova remnant scales with time and distance as
\be
\label{eq:SNr}
\begin{split}
 R_{SNR}=\frac{r_{SNR}}{r_s}&\simeq 0.1^\circ \left[\frac{5 \mathrm{kpc}}{r_s}\right]
 \left[\frac{t}{10\; \mathrm{kyr}}\right]^{0.4}\\
 &\times \left[\frac{\mathcal{E}_{SN}}{10^{51}\mathrm{erg}}\right]^{0.2} 
\left[\frac{1\;\mathrm{cm}^{-3}}{n_{ISM}}\right]^{0.2}\,,
\end{split}
\ee
where we assumed that the ISM is composed of
protons and used the relation $\rho_0=m_p n_{ISM}$.
The dependence on the supernova energy output and the density 
of the interstellar medium is 
quite mild, and the angular size of SNRs is defined, in essence, 
by distance and age.
From Eqs. \eqref{eq:SNr} and \eqref{eq:size1} it can be seen that 
the SNR radius is smaller than the radius of a gamma-ray halo 
at energies $E_\g \gtrsim 1$ GeV. 
This suggests that our analysis should be performed in this energy range
in order to avoid a possible overlap between the halos and SNRs.\footnote{
There can also be PWN, but its typical
extension $\sim 10$ pc is very small compared to that of SNRs or the gamma-ray halos we discuss. 
The results of this paper will be valid for systems which 
contain both pulsars and PWNe. }

The presence of a SNR or PWN around a pulsar
may complicate the escape of GeV particles; see Refs. \cite{Blasi:2013rva,Drury:2010am}.
However, as pointed out in these references, 
there are several reasons to expect that particle confinement does not 
necessarily take place even in the case of perfectly continuous shells, 
e.g., because of cross-field diffusion.

In principle, one  can expect that a gamma-ray
halo and the host pulsar can be offset due to the pulsar kick.
This offset is, however, quite
small for young pulsars with $T_{SD}\lesssim 10^4$ yr
and cannot exceed (see Ref. \cite{ref1} for typical kick velocities)
\be
\label{eq:offset}
\Delta R  \simeq  0.1^\circ \left[\frac{5 \;\mathrm{kpc}}{r_s}\right] \left[\frac{v}{10^3\; \mathrm{km/s}}\right]\left[\frac{T_{SD}}{10\; \mathrm{kyr}}\right]\,.
\ee
This offset is small compared to 
the angular size of the gamma-ray halo and we will neglect it in what follows.

\section{Pulsar sample}
\label{sec:sample}

In Ref. \cite{Neronov:2012kz} Neronov and Semikoz identified 
18 degree-scale extended sources 
(to be referred t as {\it N-S sources} after the authors of Ref. 
\cite{Neronov:2012kz} in what follows), 
most of which spatially
coincide with young pulsars with $T_{SD}\lesssim$ 30 kyr.
The most straightforward approach would be to directly analyze 
these sources with an extended set of  Fermi-LAT data. 
However, there are  several issues which complicate the direct analysis.
All but one (17 out of 18) of the 
{\it N-S} candidates are located very close to the Galactic plane,  $|b|<1^\circ$. 
This increases the possibility of background contamination and projection effects,
which may result in a false discovery of a halo.

Most {\it N-S} candidates either adjoin or spatially coincide with several
extended and point-like very high-energy (VHE) sources,
which makes it practically impossible to disentangle 
extended halos from the collective emission of these sources.
It should be noted that these sources may in fact be
inhomogeneities of halos themselves, and further investigation of this 
possibility is needed.
Some {\it N-S} candidates are so close to each other (e.g., sources No 4,5, and 6 from Table II of Ref. \cite{Neronov:2012kz}) that they form a single 
``cluster" that covers multiple VHE sources.
A few {\it N-S} sources can be associated 
with several pulsars, which further obscures their study.

In order to overcome these difficulties we follow 
an alternative method, which is to seek gamma-ray halos
in an independent ``cleaner" set of young pulsars.
For this purpose we singled out eight sufficiently isolated young nearby pulsars 
located quite away from the Galactic plane. 
In order to select these pulsars we used the ATNF catalogue 
\cite{Manchester:2004bp,atnf}
and imposed several restrictions on the pulsars' properties and location.
We put the following cuts on spin-down ages and distances:
\be 
\label{eq:cut1}
T_{SD}<30~\mathrm{kyr}\,,\quad r_s< 5~\mathrm{kpc}\,,
\ee
which select sufficiently nearby pulsars
whose hypothetical halos should have sizable fluxes and angular extensions, and
thus should be better distinguishable in the data.

In order to 
decrease the influence of the Galactic plane and the Galactic center, we 
chose the following range of Galactic coordinates: 
\be 
\quad 15^\circ<l<345^\circ\,, \quad |b|>1^\circ\,.
\ee
We obtained the set of pulsars listed in Table \ref{tab:1}. 
Note that we excluded the Vela pulsar which is very close ($r_s= 0.28$ kpc)
and relatively old ($T_{SD}=11.3$ kyr). 
The gamma-ray halo around this pulsar should have an angular size so large [$R_{halo}(1\;\mathrm{GeV})\sim 10^\circ$, see Eq. \eqref{eq:size1}]
that current diffuse models do not allow for its study~\cite{Vela}.

One can check that the distance from the Galactic disc is
smaller than 200 pc for all of the pulsars except PSR J0007+7303.
Thus, these pulsars are still situated in the dense part of the 
neutral hydrogen (HI) disc where there 
should be enough target material \cite{Kalberla}.
As for PSR~J0007+7303, a recent analysis suggests the average ISM density 
$n_{ISM}\sim~0.1$~cm$^{-3}$ \cite{Martin:2016uzs}, which implies that the halo around this pulsar could still 
have a sizable flux.

Before moving on, we check 
that our sample of pulsars belongs to a population
similar to that of the
pulsars listed in Table II of Ref.~\cite{Neronov:2012kz}
(they will be referred to as {\it N-S pulsars} in what follows).
We  have already imposed an upper bound on pulsar ages [Eq. \eqref{eq:cut1}]
which was suggested in Ref. \cite{Neronov:2012kz}.
In the scenario of CRs generated due 
to the pulsar rotational energy one might be interested 
in initial rotation periods and energy loss rates.  
These initial properties can be obtained 
only if the pulsar age is known independently from the spin-down, 
which is possible only in rather specific circumstances \cite{Noutsos:2013ce,Popov:2012ng};
this is why we instead focus on current periods and energy losses.

Using the ATNF database we found these quantities 
for the {\it N-S} pulsars and the pulsars from our set
(see Table \ref{tab:1} and Table \ref{tab:ns} in Appendix \ref{app:NSpuls}).
In order to prove that the selected set of pulsars
belongs to the same population as the {\it N-S} pulsars,
we perform a two - sample Kolmogorov-Smirnov (KS) test 
over the values of $P$ and $\dot E$ (for details, see Appendix \ref{app:NSpuls}).
We find very large p-values for either case,
$p_{KS}\sim 0.7$,
which implies that the {\it N-S} sample and our sample indeed have statistically indistinguishable  
distributions over $\dot E$ and $P$.

\begin{table}[h!]
\begin{tabular}{|c|c|c|c|c|c|c|c|}
\hline
 & PSRJ  & $l$ & $b$ & $r_s$, kpc & $T_{SD}$, kyr & $\dot E$, erg/s &$P$, s \\\hline
1 & J0007+7303 & $119.66 $ & $10.46$ & $1.40$ & $13.9$&$4.5\times 10^{35}$ &$0.32$\\\hline
2 & J0501+4516 & $161.55 $ & $1.95$ & $2.20$ & $15.7$& $1.2\times 10^{33}$&$5.8$\\\hline
3 & J1709-4429 & $343.10 $ & $-2.69$ & $2.60$ & $17.5$& $3.4\times 10^{36}$&$0.10$\\\hline
4 & J2229+6114& $106.65$ & $ 2.95$ & $3.00$ & $10.5$& $2.2\times 10^{36}$&$0.052$\\\hline
5 & J0205+6449 & $130.72 $ & $3.08$ & $3.20$ & $5.37$& $2.7\times 10^{37}$&$0.065$\\\hline
6 & J1357-6429 & $309.92  $ & $-2.51$ & $4.09$ & $7.31$& $3.1\times 10^{36}$&$0.17$\\\hline
7 & J0534+2200 & $184.56 $ & $-5.78$ & $2.00$ & $1.26$& $4.5\times 10^{38}$&$0.033$\\\hline
8 & J1513-5908 & $320.32 $ & $-1.16$ & $4.40$ & $1.56$& $1.7\times 10^{37}$&$0.15$\\\hline
\end{tabular}
\caption{Pulsars selected for likelihood analysis.}
\label{tab:1}
\end{table}

\begin{table}[h!]
\begin{tabular}{|c|c|c|c|c|c|c|}
\hline
 & PSRJ  & $R_{halo}(1\;\mathrm{GeV})$ & $R_{halo}(10\;\mathrm{GeV})$ & $R_{halo}(100\;\mathrm{GeV})$   \\\hline
1& J0007+7303 & $2.0^\circ $ & $3.2^\circ$ & $5.0^\circ$ \\\hline
2&J0501+4516 & $1.4^\circ $ & $2.2^\circ$ & $3.5^\circ$ \\\hline
3&J1709-4429 & $1.2^\circ $ & $2.0^\circ$ & $3.1^\circ$ \\\hline
4&J2229+6114& $0.8^\circ$ & $ 1.3^\circ$ & $2.1^\circ$ \\\hline
5&J0205+6449 & $0.6^\circ $ & $0.9^\circ$ & $1.4^\circ$ \\\hline
6&J1357-6429 & $0.5^\circ  $ & $0.8^\circ$ & $1.3^\circ$ \\\hline
7&J0534+2200 & $0.4^\circ  $ & $0.7^\circ$ & $1.1^\circ$ \\\hline
8&J1513-5908 & $0.2^\circ $ & $0.4^\circ$ & $0.6^\circ$ \\\hline
\end{tabular}
\caption{Theoretical expectations for the sizes of the gamma-ray
halos at different energies.}
\label{tab:halos}
\end{table}

The sizes of the halos around selected pulsars 
are computed at different energies using Eq.\eqref{eq:size1} 
and listed in Table~\ref{tab:halos}.
Comparing Tables~\ref{tab:1} and~\ref{tab:halos},
one may notice that the sample of pulsars we selected is not totally 
homogeneous
with respect to pulsar ages, spin-down luminosities, and the sizes of halos.
There is a subset of very young pulsars with $T_{SD}<10$ kyr and large energy
losses,
\[
10^{36} \;\mathrm{erg/s} \; \lesssim \dot E \lesssim \; 10^{38} \; \mathrm{erg/s}\,,
\]
which includes PSR J0205+6449, PSR J1357-6429, Crab (PSR J0534+2200), and PSR J1513-5908.
The halos around these pulsars
have quite small angular extension (see Table \ref{tab:halos}), which is why
we will dub them {\it pulsars with compact halos} in what follows. 
For energies $E_{\gamma} \simeq 1-10$~GeV the sizes of their halos 
appear to be roughly equal to the  LAT point spread function (PSF) in this range,
\be 
R_{halo}({1-10\;\mathrm{GeV}})\sim R_{\mathrm{PSF}}  ({1-10\;\mathrm{GeV}})\sim 0.5^\circ\,.
\ee
A broad PSF worsens the localization capability and  
implies
that halo photons from the energy bin $1-10$ GeV have less statistical
significance. Thus, the data in this energy range are less
sensitive to compact gamma-ray halos.
On the other hand, from 1 to 30~GeV 
the PSF falls from $0.8^\circ$ down to $0.1^\circ$ \cite{Acero:2015hja} and stays nearly constant at 
higher energies.
On the contrary, the halo 
size increases according to Eq.\eqref{eq:size1}, 
which facilitates the detection of halos by Fermi-LAT at energies $E_\g>10$ GeV.

The remaining 4
pulsars from our set (PSR J0007+7303, PSR J0501+4516, PSR J1709-4429, and PSR J2229+6114) form a subsample of relatively old (10 kyr $<T_{SD}<$ 30 kyr)
pulsars with moderate energy losses,
\[
10^{33}\; \mathrm{erg/s}\; \lesssim \dot E \lesssim \; 10^{36} \;\mathrm{erg/s}\,.
\]
Looking at Table \ref{tab:halos}, one can make sure that the size of the LAT PSF
is smaller than the angular extension of gamma-ray halos around these pulsars above 1 GeV.
This means that the halos should be better observed in the energy bin 1-10 GeV
where one can expect the largest flux.
We will refer to this subset of pulsars as {\it pulsars with large halos}.

\section{Data analysis}
\label{sec:method}

In our analysis we use the Fermi-LAT data collected 
during 361 weeks from August 04, 2008 (MET=239557418s) to July 6, 2015 (MET=457859500). 
We use the Fermi science tools\footnote{http://fermi.gsfc.nasa.gov/ssc/data/analysis/} (version v10r0p5), including the Pass 8 reconstruction (P8R2\_SOURCE\_V6).
We have selected events belonging to the ``SOURCE" class in order to have a reasonable number 
of events of good quality. 
When processing the data, we strictly followed the routine described in Ref.~\cite{p8manual},
which included the zenith angle cut of 90$^\circ$. 
Moreover, data 
collected while the observatory was passing
across the South Atlantic Anomaly
were not taken into consideration.

In order to trace the variation of the halo size with energy as predicted by
Eq. \eqref{eq:size1} we split the data into three different energy bins, 
100-500 GeV, 10-100 GeV, and 1-10 GeV, and 
analyze them separately.
The selected events with energies $1$ GeV$\leq E_\g\leq 500$ GeV
have relatively small PSF values, $R_{\mathrm{PSF}}<0.8^\circ$ which allows 
us to use smaller regions of interest (RoIs). 
In practice, we take a circle of radius $10^\circ$ around each pulsar.
The data are analyzed using the binned likelihood approach 
implemented in the \textit{gtlike} utility, in which 
two model hypothesis are compared by their maximal likelihoods 
with respect to the observed photon
distribution. The null hypothesis does not include new sources compared to 
the 3FGL catalogue \cite{Acero:2015hja}, 
while the alternative hypothesis assumes a 
halo around a selected pulsar added into the list of sources of the null hypothesis. 

The null source model for each pulsar 
includes all of the sources from the 3FGL catalogue taken within 
a 10$^\circ$ radius around the selected pulsar, the corresponding 
galactic interstellar emission model \textit{gll\_iem\_v06.fits}, 
and the isotropic spectral template \textit{iso\_P8R2\_SOURCE\_V6\_v06.txt}.
We use the spectral models from 3FGL and keep spectral parameters free
for all sources within the RoI in the likelihood optimization procedure.

Now we discuss how the sources associated with the pulsars of interest are
modeled in the 3FGL catalogue and hence in our input models.
The pulsar PSR J0501+4516 (No 2 in Table~\ref{tab:1})
has a rather long period and low spin-down luminosity; 
it does not have any gamma-ray counterpart in 3FGL and hence is absent in our model.
The pulsar No 8 (PSR J1513-5908)
is modeled in 3FGL as a point-like source
while its PWN MSH 15-52 is modeled as a separate extended source of size $0.04^\circ\times 0.11^\circ$. 
A complex spectrum of the Crab source (No 7 in our list)
has been reconstructed in 3FGL by means of
three different components 
\cite{Acero:2015hja}:
the gamma-pulsar with an exponential cutoff, 
a soft power-law synchrotron emission of the Crab PWN 
and a hard power-law inverse Compton emission of this PWN. 
The sources associated with the pulsars No 1,3,4, and 5 are modeled with the 
exponential cutoff power-law spectrum typical for pulsars. 
The source associated with the pulsar No 6 is modeled as a point source 
with a power-law spectrum,
which suggests that it corresponds to a pulsar-PWN system.

For the alternative hypothesis,
on top of the LAT sources discussed above
we have added spatial templates centered at the
pulsars' coordinates taken from the ATNF database. 
For the extended halos we use the simplest spatial models: uniformly bright circles of different radii (from 0 to 5 degrees with a 0.1 degree step).

We use the simplest spatial model of a uniformly bright disc
to remain maximally model independent.
This is obviously an oversimplification, since it is expected that 
halos can have more complex morphology \cite{Giacinti:2012ar,Giacinti:2013uya,Kachelriess:2015oua}.
However, recent studies
imply that the use of the simplest templates is quite robust:
it does not drastically alter  
the statistical significance of halo detection along with
best-fit values of fluxes and spectral indices~\cite{Pshirkov:2016qhu}.

The spectrum of the gamma-ray halos
was taken as a simple power law,
\be
\label{eq:spectrum}
\frac{dN}{dE}=N_0\left(\frac{E}{E_0}\right)^{-\Gamma} \,,
\ee
where the normalization factor $N_0$ and the spectral index $\Gamma$
are allowed to vary during the likelihood analysis, while the energy $E_0$
is fixed at 1 GeV.

The evidence of the detection of extended gamma-ray emission around the pulsars is evaluated in terms of the likelihood ratio test statistic ($TS$):
\be 
TS = -2\ln \frac{L_{max,0}}{L_{max,1}}
\ee
where $L_{max,0}$ and $L_{max,1}$ are the maximum likelihood values obtained when fitting the observed data using the null and alternative hypothesis, respectively. 
Note that $\sqrt{TS}$ is approximately equivalent 
to the source detection significance.

\section{Simulations}
\label{sec:simul}

Before analyzing the real Fermi data, 
in order to estimate the sensitivity of our method to gamma-ray halos
we apply our method to the simulated event sets that include
the halo in the source model. 

Our sample of pulsars is divided into two subsets,
which have different  properties and are expected to be pretty
different from the observational point of view. 
In order to understand these differences we 
chose to simulate one pulsar from each subset.
We chose the pulsar No~1 (PSR J0007+7303) to represent the pulsars with large halos and
the pulsar No~8 (PSR J1513-5908) to represent the pulsars with compact halos.
These pulsars are bracketing cases for our set. 
The pulsar PSR J0007+7303
is very close, located far away from the 
Galactic plane and its halo should have the biggest 
angular extension among other pulsars. 
On the contrary, the pulsar PSR J1513-5908 is the farthest away,
located close to the Galactic plane, and has the smallest angular size of a hypothetical halo.

In this section we briefly report the main outcome of our simulations
performed with the use of the {\it gtobssim} utility. 
The details may be found in Appendix \ref{app:simul}.
The simulated events are processed using the 
{\it gtlike} utility analogous to the real data (see Sec. \ref{sec:method}).

\begin{figure}[h!]
\includegraphics[width=0.5\textwidth]{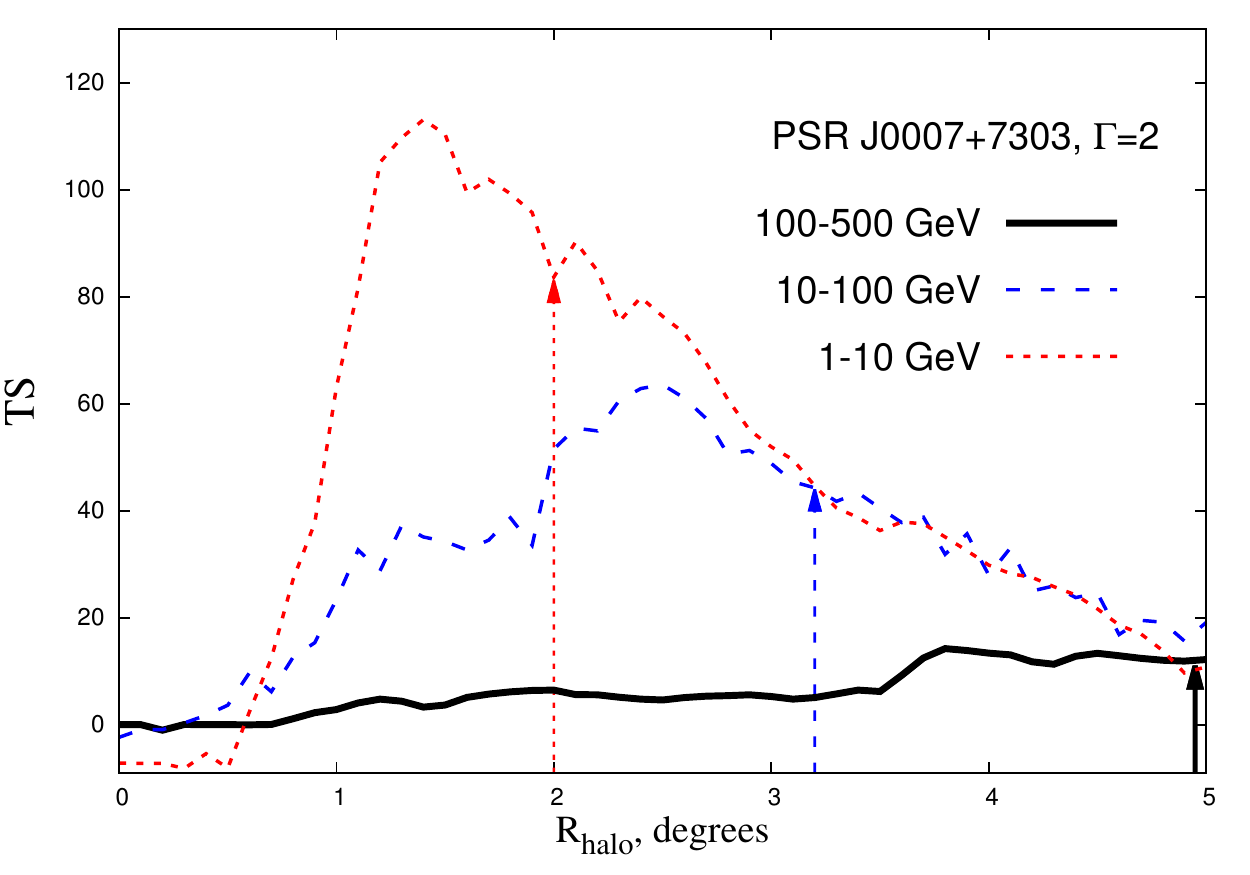}
\includegraphics[width=0.5\textwidth]{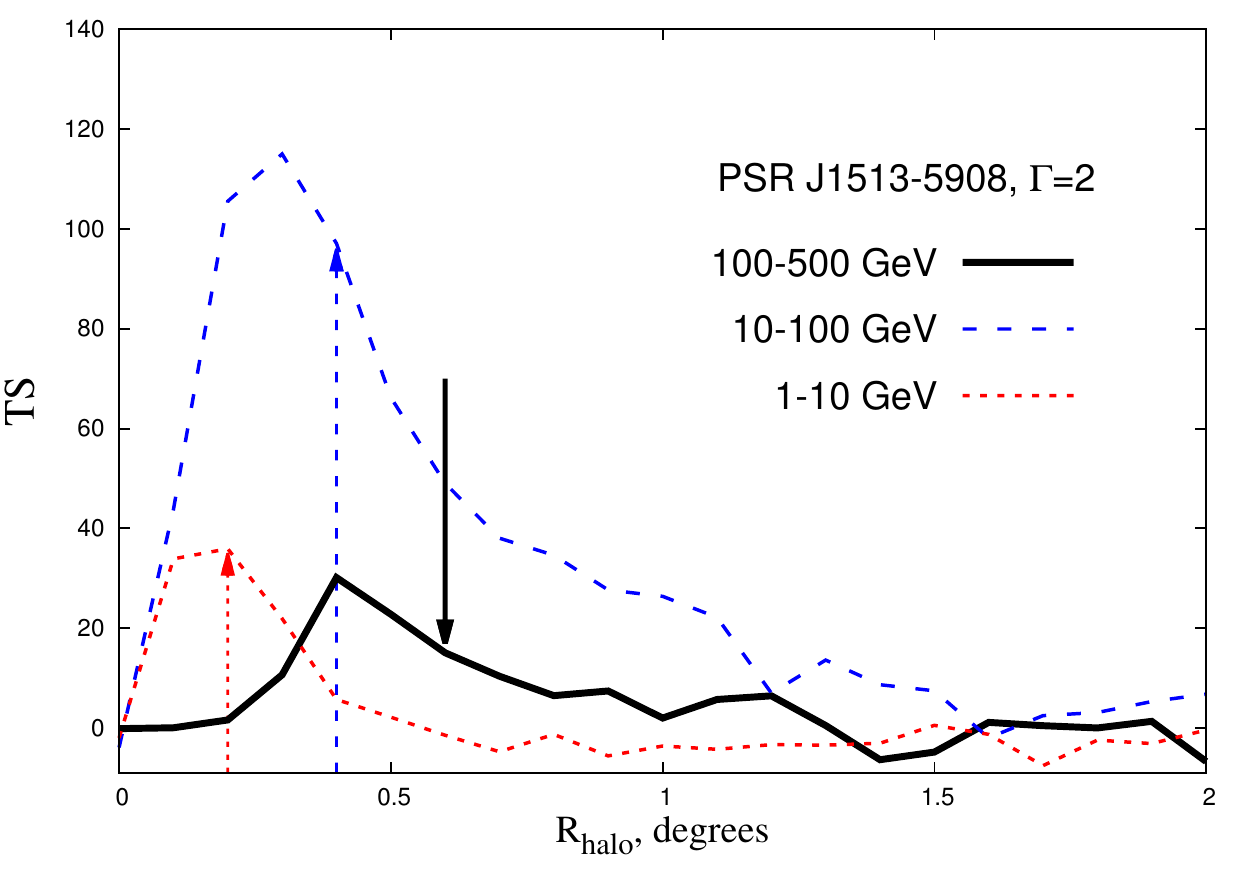}
\caption{\label{fig:sim_faint}
$TS(R_{halo})$ curves for 
the simulated {\it faint} gamma-ray halos around the pulsar PSR J0007+7303 (upper panel) and
PSR J1513-5908 (lower panel); see Appendix \ref{app:simul}.
The results of the analysis in different energy bands
are shown as a black solid line for 100-500 GeV,  
a blue dashed line for 10-100 GeV, and a red dotted line for 1-10 GeV. 
Vertical arrows show the sizes of the halos 
that were used in the simulations.
}
\end{figure}

For either pulsar we simulate two different types of gamma-ray
halos. We call them {\it bright} and {\it faint} halos.
For the bright halos 
we assume the fluxes as reported by Ref. \cite{Neronov:2012kz}
[of order Eq.\eqref{eq:flux} in the energy bin 100-500 GeV,
or, equivalently, the overall luminosities of order Eq.~\eqref{eq:nslum}].
Our results imply that in this case 
gamma-ray halos will be detectable 
around all of the pulsars from our set
in all three energy bins at quite high statistical significance.

In the case of faint halos 
we follow a more phenomenological approach.
For either pulsar we seek the flux which produces the signal 
with significance $TS~\sim~100$ in at least one of the energy bins.
As anticipated,  the sensitivity appears to be quite different
for the two subpopulations of pulsars
(see Fig.~\ref{fig:sim_faint}).

\begin{figure}[h!]
\includegraphics[width=0.5\textwidth]{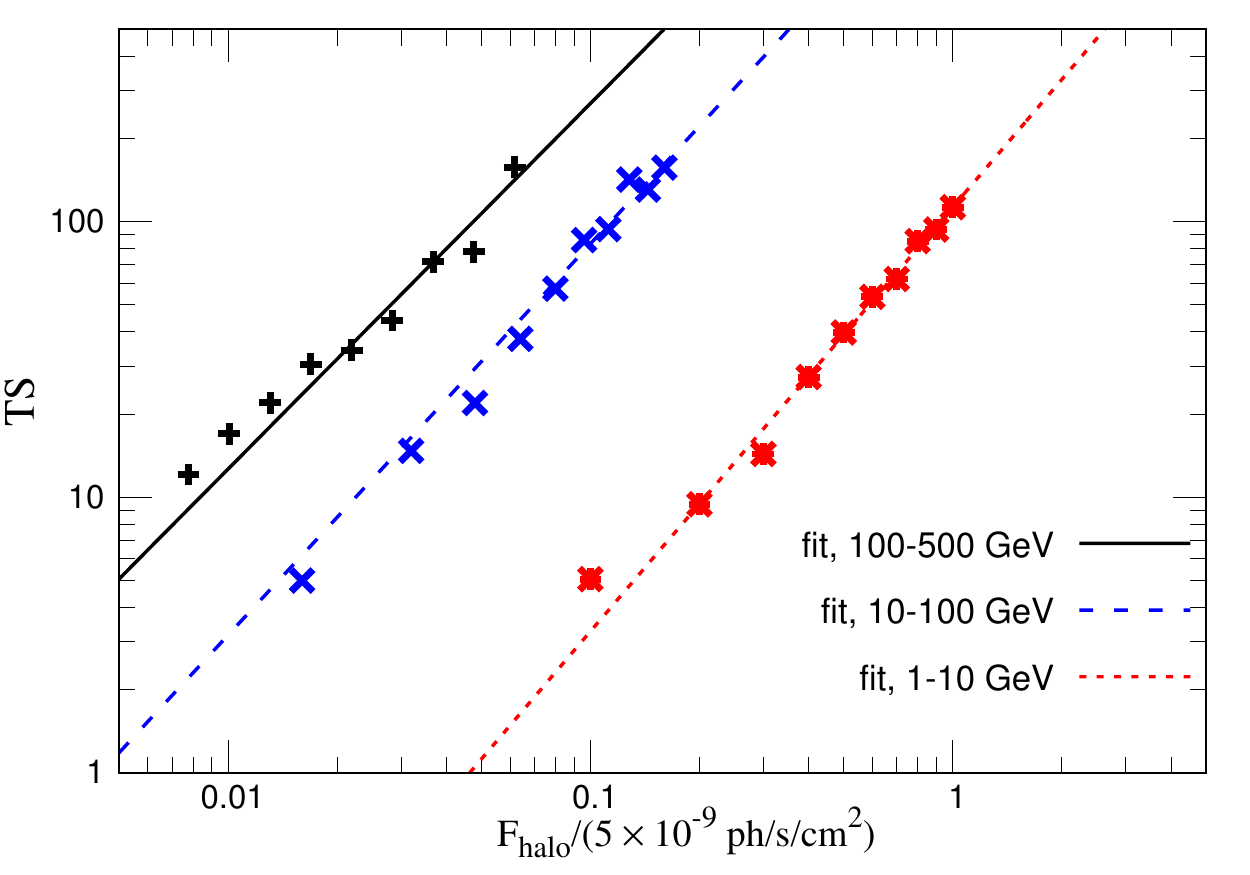}
\caption{\label{fig:scal_m}
The $TS(F_{halo})$ dependence retrieved from simulations. 
Data points are taken from maxima of measured TS curves, 
and the lines represent best fits given in Eq.~\eqref{eq:tscta_m}.
The events used to produce this plot are generated for a halo around PSR J0007+7303 with
$\Gamma=2$; see App.~\ref{app:tss} for more details. 
}
\end{figure}

By scanning over different values of fluxes we find that
in the case of pulsars with large halos
our method is most sensitive to their fluxes
in the energy bin 1-10 GeV 
(see the upper panel of Fig. \ref{fig:sim_faint}), in which the flux
$F^{1-10\;\mathrm{GeV}}\simeq 5\times 10^{-9}$ ph/cm$^{2}$s
yields a halo detection with the desired~significance.

On the contrary, the compact halos appear to be 
more easily detectable in the energy bin 
10-100 GeV 
because of the lack of resolution at 1-10 GeV (see the lower panel of Fig. \ref{fig:sim_faint}).
We find that the flux $F^{10-100\;\mathrm{GeV}}~\simeq~6~\times~10^{-10}$~ph/cm$^{2}$s
leads to a halo detection at $TS\sim 100$, 
while a detection at the same significance in the 
energy bin 1-10 GeV requires 
an order of magnitude larger flux in this bin.

We study the dependence of test statistics on the halo flux.
For that we vary the input flux and find the resulting 
values of maxima of corresponding $TS$ curves (See Fig. \ref{fig:scal_m}).
As a result, we obtain the following scaling:
\be
\label{eq:tscta_m}
\begin{split}
TS_{1-10}\simeq & 100 \left[\frac{F^{1-10\;\mathrm{GeV}}}{4.6\times 10^{-9}\;\mathrm{ph/cm}^2\mathrm{s}}\right]^{1.54}\,,\\
TS_{10-100}\simeq & 100 \left[\frac{F^{10-100\;\mathrm{GeV}}}{5.7\times 10^{-10}\;\mathrm{ph/cm}^2\mathrm{s}}\right]^{1.42}\,,\\
TS_{100-500}\simeq & 100 \left[\frac{F^{100-500\;\mathrm{GeV}}}{2.4\times 10^{-10}\;\mathrm{ph/cm}^2\mathrm{s}}\right]^{1.33}\,,
\end{split}
\ee
which holds true if the angular size of the halo is larger than the LAT PSF.
The dependence of this scaling 
on other parameters (e.g., the halo spectral index, the galactic latitude, etc.)
is found to be quite mild and cannot exceed $20\%$ for the range of interest $10\lesssim TS\lesssim 100$; see Appendix \ref{app:simul}
for more details.

Our analysis implies that the scaling in
the energy bins 10-100 GeV and 100-500 GeV given in 
Eq.~\eqref{eq:tscta_m} is a generic feature valid for any halo from 
both subpopulations. On the contrary, the scaling
in the energy bin 1-10 GeV holds only for pulsars with large halos.

\section{Results}
\label{sec:results}

In this section we report the 
results of searches for the gamma-ray halos in the 7-year Fermi-LAT data.
We discuss separately the outcome of our study for either subpopulation of
pulsars from Table \ref{tab:1}. 

\subsection{Pulsars with large halos}

As discussed in the previous section, the pulsars with large halos
are the best targets for our method because it is most sensitive 
to halo fluxes in the energy bin 1-10 GeV where one expects the strongest signal.
That is why, if CR halos exist, they are likely to be detected
in this set of pulsars.

\begin{figure}[h]
\includegraphics[width=0.5\textwidth]{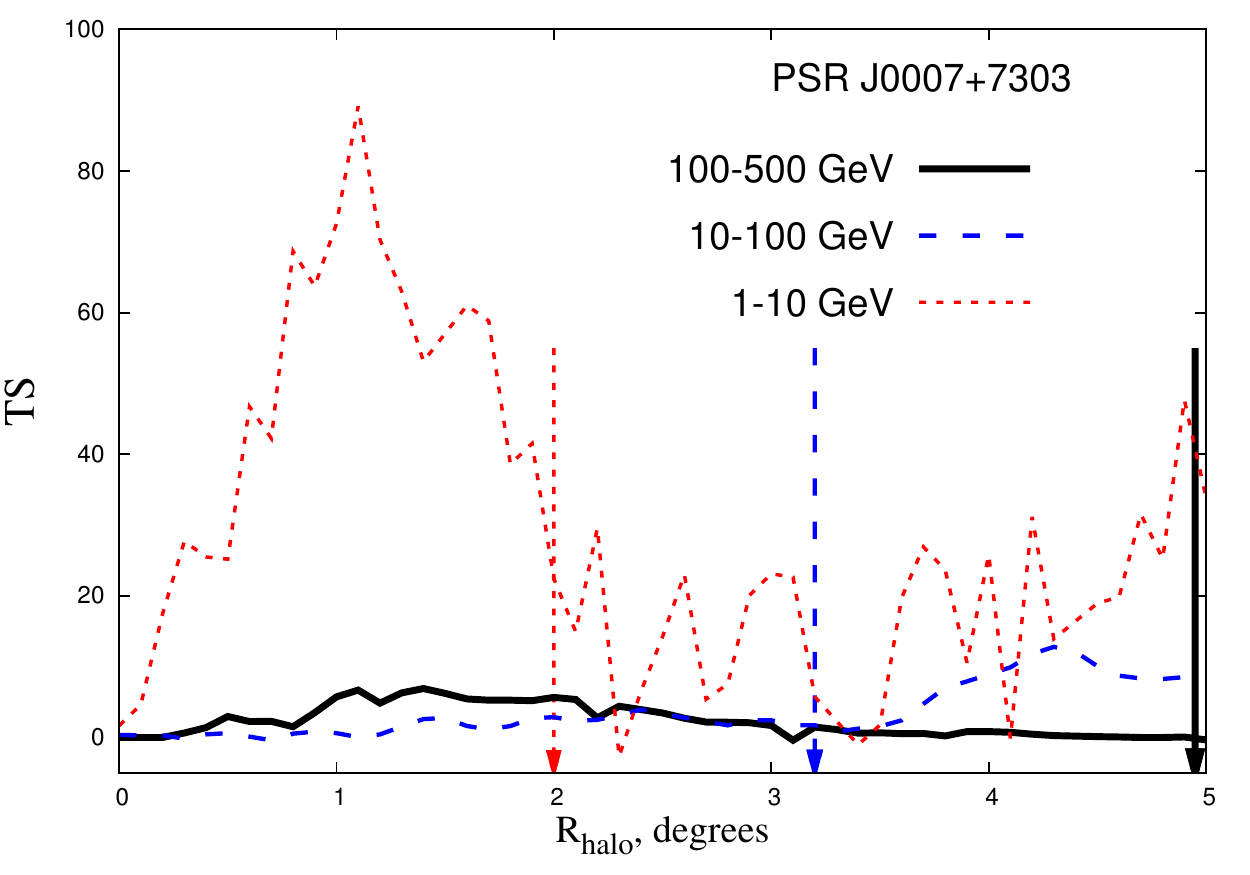}
\includegraphics[width=0.5\textwidth]{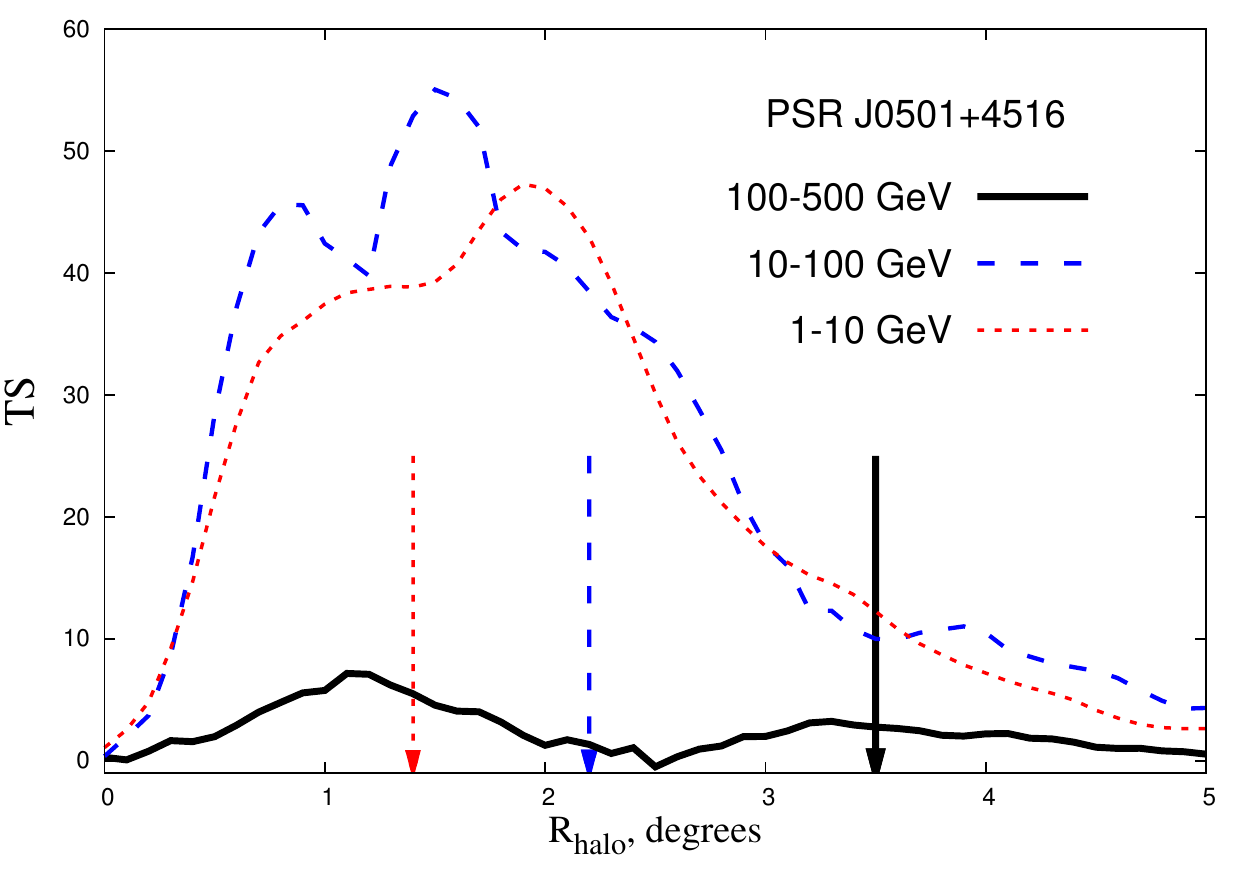}
\caption{\label{fig2}
$TS(R_{halo})$ curves for PSR J0007+7303 (upper panel) and PSR J0501+4516 (lower panel). The results of the analysis in different energy bands
are shown as a black solid line for 100-500 GeV,  
a blue dashed line for 10-100 GeV, and a red dotted line for 1-10 GeV. 
Vertical arrows show the sizes of the halos
that are expected from the estimate \eqref{eq:size1}.
}
\end{figure}

\textbf{1)} 
The analysis of the pulsar PSR J0007+7303 
reveals a degree-scale excess with $TS=89$ ($\sim 9.5 \sigma$) in the energy range 1-10 GeV.
This signal (see the upper panel of Fig. \ref{fig2}) 
can be compared to the simulations; see Figs.~\ref{fig:sim_faint} and~\ref{fig:sim_007_faint}.
Given some common features, one can interpret this excess
as a gamma-ray halo produced by CRs. 
The small excess with $TS=13$ at $R_{halo}\approx 4.2^\circ$ in the energy
bin 10-100 GeV can also be interpreted as a counterpart of the signal seen in the bin 1-10 GeV. 
The data in the energy bin 1-10 GeV yield the following fit for the flux and
spectral index at $R_{halo}=1.1^\circ$:
\be
\label{eq:ctafit}
\begin{split}
& F^{1-10\;\mathrm{GeV}}=(3.53\pm 0.23)\times 10^{-9}\;
\mathrm{photons}/\mathrm{cm}^2\mathrm{s}\,,\\
&\Gamma =2.798\pm 0.081\,.
\end{split} 
\ee

On the other hand, this excess 
may be associated with SNR CTA1 (G119.5+10.2) or a PWN.
The extended gamma-ray emission (0.1-100 GeV) of size 
$0.6^\circ\pm 0.3^\circ$ at the position of SNR 
CTA1 was discovered in the energy band 0.1-100 GeV in Ref.~\cite{Abdo:2011ce}.
Moreover, the extended TeV emission of size $0.3^\circ\times 0.24^\circ$
in the vicinity of 
PSR J0007+7303 was reported by VERITAS \cite{Aliu:2012uj}. 
This emission was suggested to be associated with a PWN,
which is supported by observations in other energy bands 
\cite{Halpern:2004qg,Mignani:2013js,Martin:2016uzs}. 
Note, however, that the extension of this emission is much smaller 
compared to the size of excess that we found.
Thus, the presence of a compact PWN does not exclude
the interpretation of the degree-scale gamma-ray emission
as a CR halo. 

Let us estimate the total luminosity of this halo.
Using the best fit \eqref{eq:ctafit} and assuming 
that the halo spectrum has the same power-law index 
at energies above 10 GeV,
one can find the total flux
(notice that we switched to the $\mathrm{erg}/\mathrm{cm}^2\mathrm{s}$ units),
\be
\label{eq:bound0}
F^{E_\g\geq 1\;\mathrm{GeV}}\simeq 1.3\times 10^{-11} \; \mathrm{erg}/\mathrm{cm}^2\mathrm{s}\,,
\ee
which yields the luminosity
\be
\label{eq:lum007}
 L_\g^{E_\g\geq 1\;\mathrm{GeV}}
\simeq 3.0\times 10^{33}
\mathrm{erg}/\mathrm{s}\,.
\ee

\begin{figure}
\includegraphics[width=0.50\textwidth]{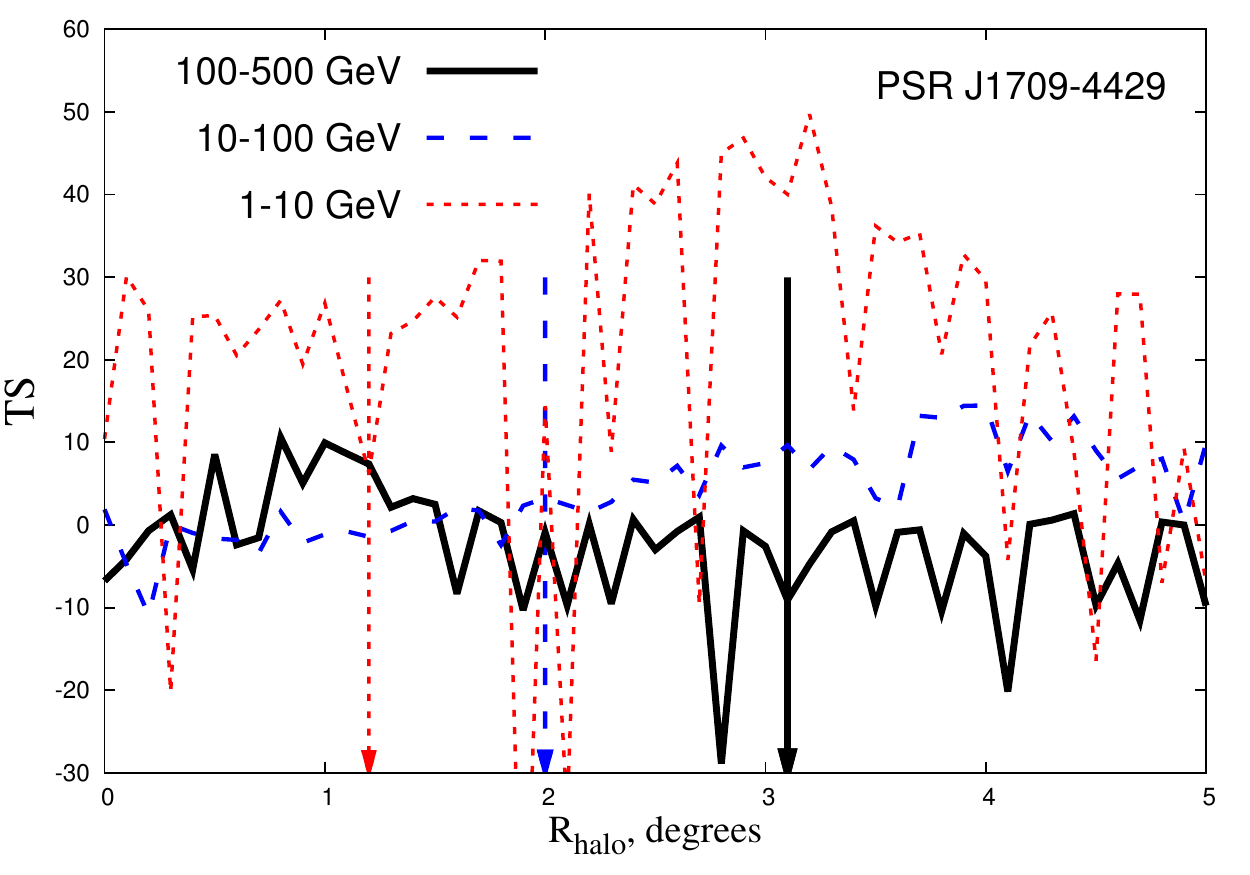}
\includegraphics[width=0.50\textwidth]{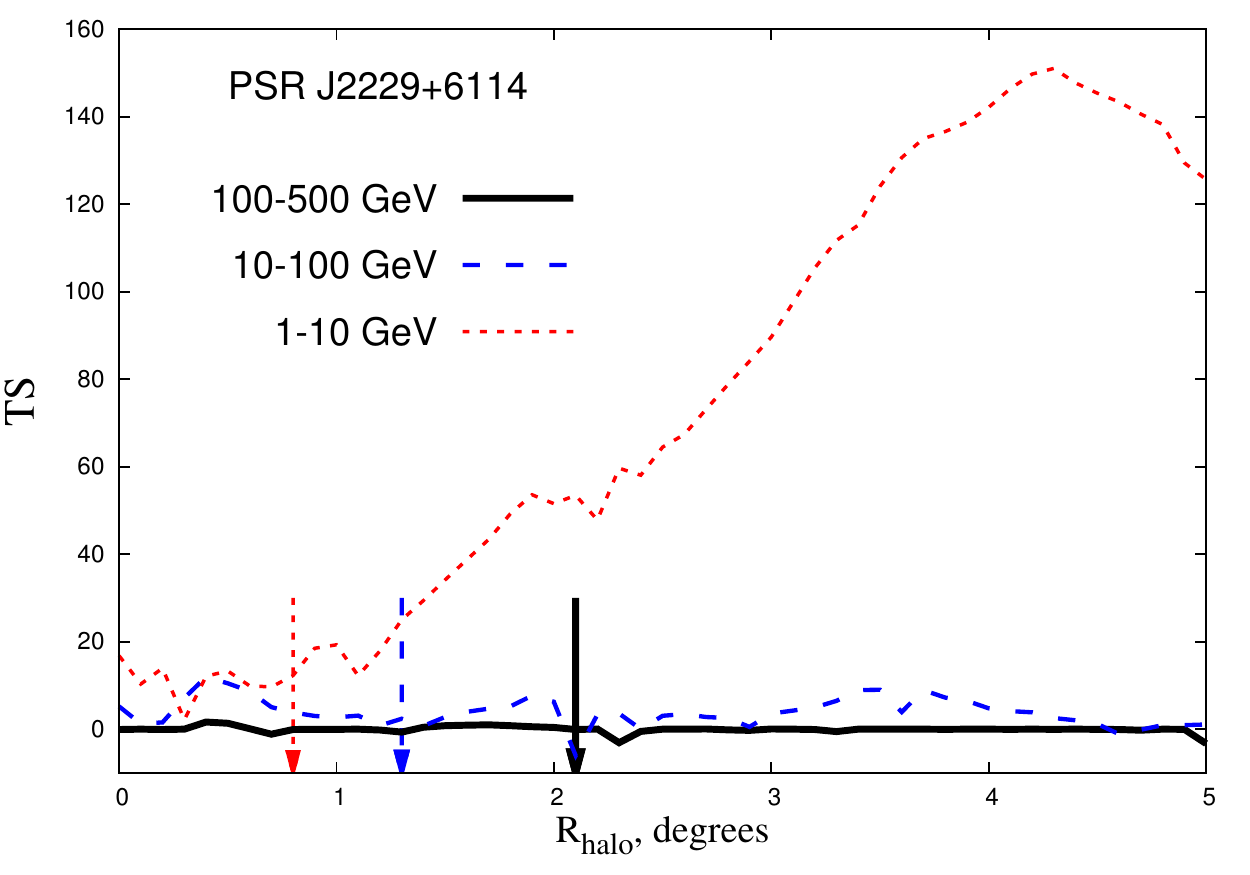}
\caption{
\label{fig4}
$TS(R_{halo})$ curves for PSR J1709-4429 (upper panel) and  PSR J2229+6114 (lower panel). 
Vertical arrows show the sizes of the halos
that are expected from the estimate \eqref{eq:size1}.
}
\end{figure}

Since a part of the signal we are looking for may be already absorbed in 3FGL sources, 
we also keep their spectral parameters free
during our analysis. 
Their values for PSR J0007+7303 can be found in Appendix~\ref{app:3fgl}.

\textbf{2)} 
When analyzing the region near the pulsar  PSR J0501+4516 
we find an excess in the energy bands 10-100 GeV and 1-10 GeV
at statistical significance $TS\simeq 45$ and $TS\simeq 55$, respectively,
and the corresponding halo size is roughly $1.5^\circ$ (see the lower panel of Fig. \ref{fig2}).

In fact, the region of interest has been studied in detail in Ref.~\cite{Araya:2014kra}.
This study has revealed the presence of a significantly extended ($R= 1.2^\circ\pm 0.3^\circ$)
gamma-ray source at the position 
of SNR HB9 [SNR G160.4+02.8, $(l,b)=(160.4^\circ,2.75^\circ)$]. 
With the new Fermi-LAT data we rediscovered this source,
but its interpretation as a CR halo
does not seem to be plausible.
The angular size of a CR halo is expected to increase with energy,
while the size of the observed emission stays nearly similar in both energy bins,
which suggests that this emission may be attributable to a SNR.
Because of this source, it is practically impossible to extract the signal
from a hypothetical gamma-ray halo. 
One can, however, place a trivial bound from the fact that 
the halo flux in the energy bin 1-10 GeV is smaller than the total observed flux. 
Using the best fit for the extended emission at $R_{halo}=1.9^\circ$, we get
\be
F^{1-10\;\mathrm{GeV}}<F_{tot}^{1-10\;\mathrm{GeV}}=(2.82\pm 0.42)\times 10^{-9}\;\mathrm{ph}/\mathrm{cm}^2\mathrm{s}\,. 
\ee
Assuming the spectral index of a halo above 10 GeV $\Gamma=2.4$,  
this yields the following bounds on the overall flux and 
luminosity of the halo above 1 GeV:
\be
\label{eq:const0501}
\begin{split}
&F^{E_\g \geq 1\;\mathrm{GeV}}<1.7\times 10^{-11}\;\mathrm{erg}/\mathrm{cm}^2\mathrm{s}\\
 &L_\g^{E_\g \geq 1\;\mathrm{GeV}}<9.3 \times 10^{33}\; \mathrm{erg}/\mathrm{s}\,.
\end{split}
\ee
The case of $\Gamma=2$ will be discussed below.

\textbf{3)}
The analysis of the pulsar PSR J1709-4429 did not reveal any sign
of extended emission (see the lower panel of Fig. \ref{fig4}). 
The data give sawtoothed $TS$ curves
without any smooth peaks in all three energy ranges.

Note that an extended emission of size $R= 0.29^\circ\pm 0.04^\circ$ above $100$ GeV 
around this pulsar 
has been detected by the HESS Collaboration
\cite{Collaboration:2011wva}.
In 3FGL the corresponding source is modeled 
as a point source, and our analysis shows that 
the extension seen by HESS is not resolved in the Fermi-LAT data.
In any case, the angular extension of the HESS excess 
is very small and cannot be interpreted as a gamma-ray halo.

The $TS$ curve for this pulsar in the energy bin 1-10 GeV lies systematically below the line $TS= 50$ 
for all halo radii.
The scaling of $TS$ with the halo flux \eqref{eq:tscta_m} implies that 
the nonobservation of a halo at this significance can be translated into a bound on the corresponding flux,
\be 
TS_{1-10}<50 \quad \Rightarrow \quad  F^{1-10\;\mathrm{GeV}}< 3.0\times 10^{-9} 
\;\mathrm{ph}/\mathrm{cm}^2\mathrm{s}\,.
\ee
Assuming the spectral index $\Gamma=2.4$, 
this gives the following constraints:
 \be
 \label{eq:const1709}
 \begin{split}
 &F^{E_\g \geq 1\;\mathrm{GeV}}<1.7\times 10^{-11}\;\mathrm{erg}/\mathrm{cm}^2\mathrm{s}\,,\\
 &L_\g^{E_\g \geq 1\;\mathrm{GeV}}<1.4 \times 10^{34}\; \mathrm{erg}/\mathrm{s}\,.
 \end{split} 
 \ee

\textbf{4)}
The results of our study for the pulsar PSR J2229+6114 are shown in 
the lower panel of Fig. \ref{fig4}.
The $TS$ curve above 100 GeV is almost flat
and coincides with the $TS=0$ axis.
The $TS$ curve in the range 10-100 GeV has a small insignificant peak at 
$R_{halo}\approx 0.5^\circ$ with the value $TS\approx 10$.
In the range 1-10 GeV the $TS$ curve features a slight enhancement 
over the range $R_{halo}\approx 0.5-1^\circ$ with $TS\sim 10$ and a
significant peak at $R_{halo}\approx 4.5^\circ$
with $TS\sim 150$.

The extended emission of size $0.5^\circ$ seen in the range 
10-100 GeV likely
corresponds to PWN G106.65+2.96 
(associated with SNR G106.3+2.7; see Refs.~\cite{Kothes:2001px,Acciari:2009zz}),
whose counterpart was modeled as a point source 
in the 3FGL catalogue (and hence, in our source model). 
The latter accounts for the marginal improvement 
of $TS$
when adding an extended template to the source model.
We conclude that for a given pulsar
the data do not show any evidence for extended 
emission which can be attributed to a gamma-ray halo.

The emission observed at $R\approx 4.5^\circ$ 
might originate from the Galactic plane.
In any case, such a large angular 
separation
(which would correspond to $\sim 200$ pc 
if projected at the pulsar's distance)
implies that this emission is not related to 
the pulsar of interest.

Analogous to the previous pulsar, the fact that the $TS$ curve for PSR J2229+6114
in the energy bin 1-10 GeV lies below the line $TS= 30$ 
at halo sizes $R_{halo}\lesssim 1.5^\circ$ can be used to put a bound on the halo luminosity,
\be 
TS_{1-10}<30 \quad \Rightarrow \quad  F^{1-10\;\mathrm{GeV}}< 2.0\times 10^{-9} 
\;\mathrm{ph}/\mathrm{cm}^2\mathrm{s}\,.
\ee
Assuming the spectral index $\Gamma=2.4$, 
this implies
 \be
 \label{eq:const2229}
 \begin{split}
 &F^{E_\g \geq 1\;\mathrm{GeV}}<1.1\times 10^{-11}\;\mathrm{erg}/\mathrm{cm}^2\mathrm{s}\\
 &L_\g^{E_\g \geq 1\;\mathrm{GeV}}<1.2 \times 10^{34} \mathrm{erg}/\mathrm{s}\,.
 \end{split} 
 \ee

An important step in deriving the constraints on halo luminosities 
was the choice of the spectral 
index $\Gamma~=~2.4$. 
The constraints, essentially, do not change under the assumption of harder spectra.
Indeed, in this case one can constrain the halo luminosity by using the signal in the energy bin 10-100 GeV [see Eq.~\eqref{eq:tscta_m}].
For instance, 
having assumed the slope with $\Gamma=2$ one can derive the following constraint 
for the PSR J2229+6114 case:
\be
\label{eq:bin2large}
\begin{split}
&F^{1 \mathrm{GeV}\leq E_\g \leq 500\;\mathrm{GeV}}<1.8\times 10^{-11}\;\mathrm{erg}/\mathrm{cm}^2\mathrm{s}\\
 &L_\g^{1 \mathrm{GeV}\leq E_\g \leq 500\;\mathrm{GeV}}<2.0 \times 10^{34} \mathrm{erg}/\mathrm{s}\,,
\end{split}
\ee
which stays, essentially, at the same level as Eq.\eqref{eq:const2229}.
Note that the difference between Eq.\eqref{eq:const2229} and Eq.\eqref{eq:bin2large} 
can be used as an estimate for an error introduced by
spectra extrapolations. We see that it brings $\sim 30\%$ uncertainty.

\subsection{Pulsars with compact halos}

\begin{figure}
\includegraphics[width=0.50\textwidth]{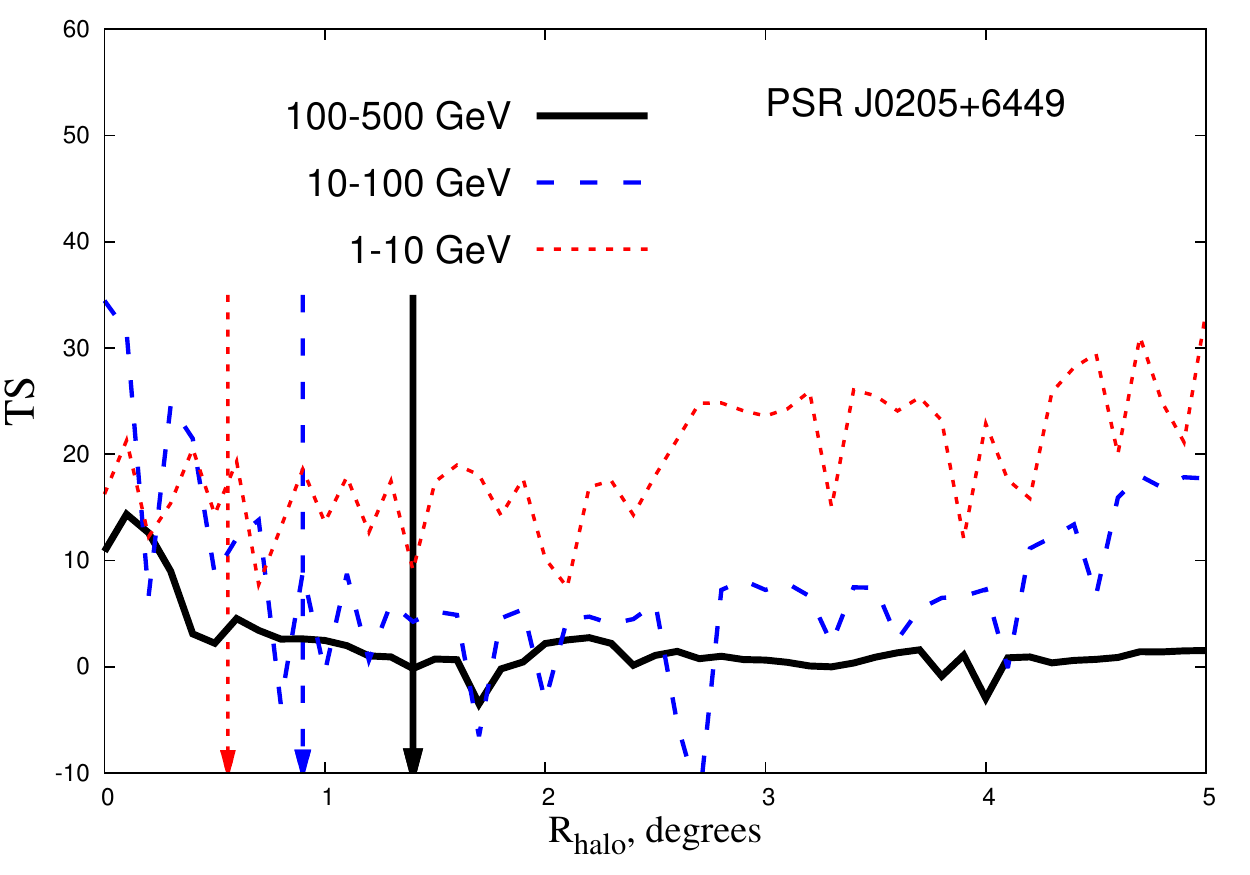}
\includegraphics[width=0.50\textwidth]{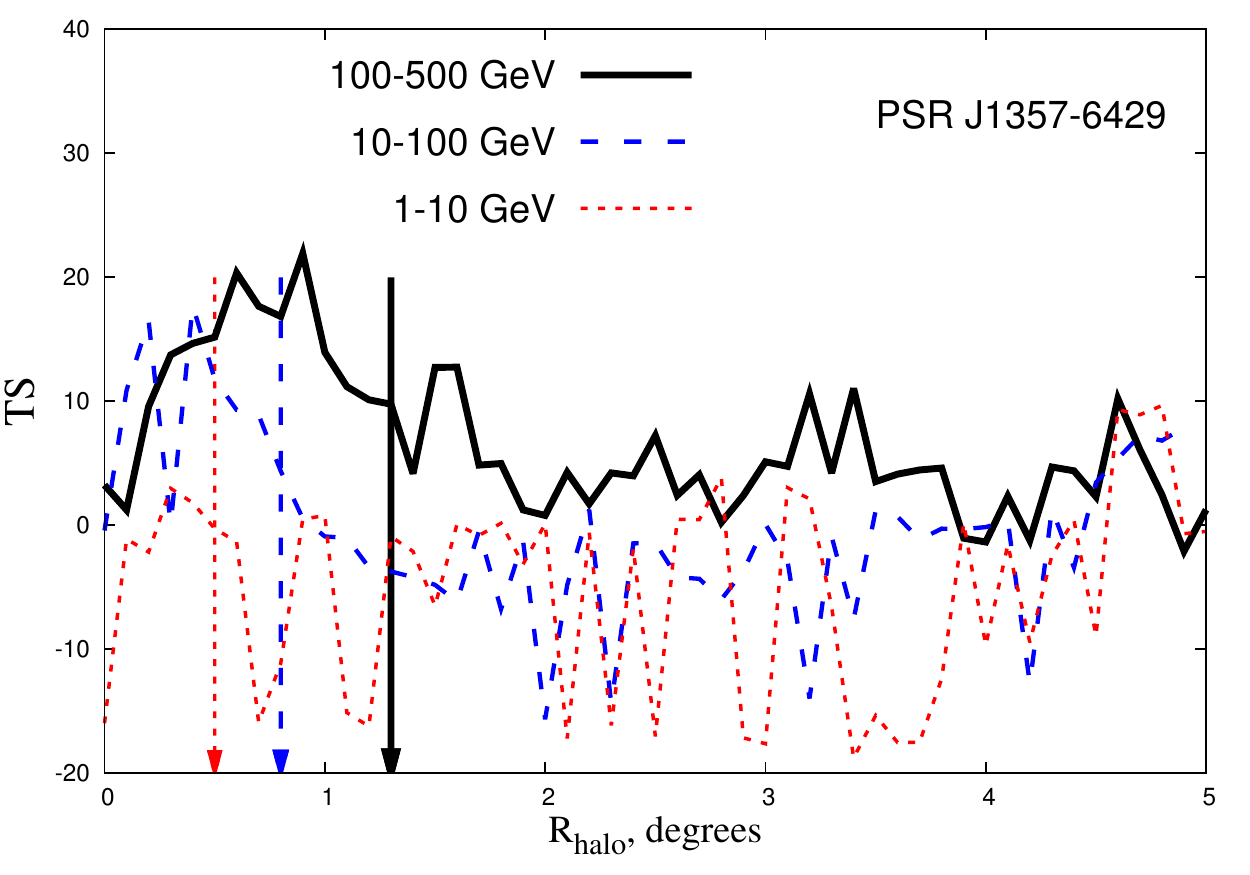}
\caption{
\label{fig3}
$TS(R_{halo})$ curves for PSR J0205+6449 (upper panel) and PSR J1357-6429 (lower panel). 
Vertical arrows show the sizes of the halos 
that are expected from the estimate \eqref{eq:size1}.
}
\end{figure}

\textbf{5)} For the pulsar PSR J0205+6449 the data show no evidence of extended emission in 
all three energy bands (see the upper panel of Fig. \ref{fig3}). The $TS$ curves lie around zero
in the ranges 10-100 GeV and 100-500 GeV, while in the band 1-10 GeV
the $TS$ curve oscillates around a constant value $TS\approx 20$, which 
suggests that this offset resulted from inaccuracies in the 
background modeling.

\textbf{6)} 
The $TS$ curve for the pulsar PSR J1357-6429 (see the lower panel of Fig. \ref{fig3}) 
is quite jagged above 100 GeV 
and has a wide peak with two small spikes 
at $R_{halo}\approx 0.6^\circ$ with the significance 
$TS\approx 20$. 
In the energy band 10-100 GeV the $TS$ curve lies near the zero axis
for $R_{halo}\gtrsim 1^\circ$ and features a small excess at $R_{halo}\approx 0.3^\circ$
with the maximum values $TS \approx 16$.
The $TS$ curve for 1-10 GeV lies around zero for almost all halo radii.

The excess above 100 GeV is likely associated with the extended 
HESS J1356-645 source 
\cite{Chang:2011pz,Abramowski:2011ia}, whose counterpart in the 3FGL 
catalogue (3FGL J1356.6-6428) 
is modeled as a point source. 
This explains the marginal improvement of the $TS$
achieved by adding an extended template of size $R_{halo}\sim 0.5^\circ$.
The combination of
radio, x-ray, and gamma-ray observations 
indicates that this source is a PWN,
whose gamma emission has a leptonic origin \cite{Abramowski:2011ia}.

On the other hand, we see that adding extended templates
does not improve the $TS$ significantly in the 1-10 GeV and 10-100 GeV bins, 
which indicates the absence of any evidence of a  gamma-ray halo, at least at the present level of sensitivity.

\begin{figure}
\includegraphics[width=0.50\textwidth]{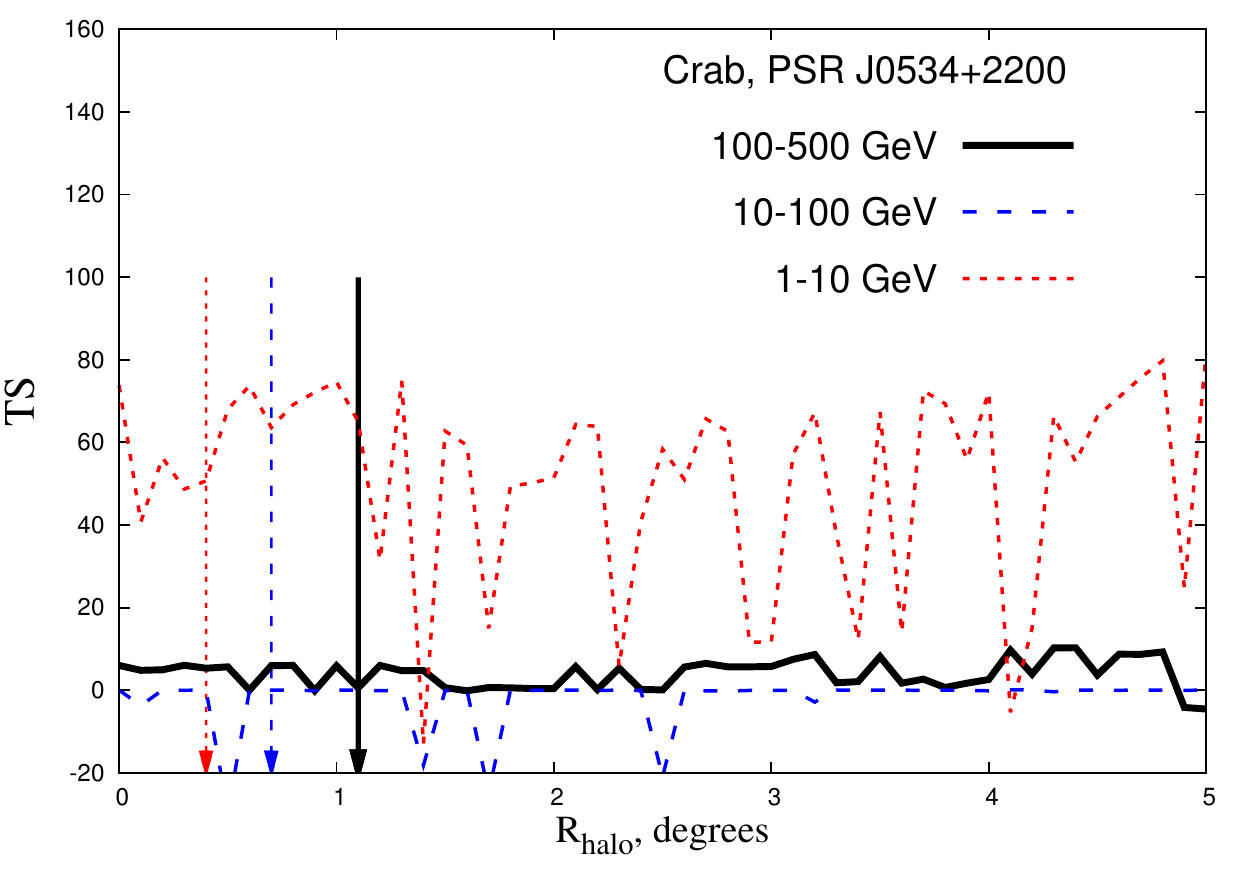}
\includegraphics[width=0.50\textwidth]{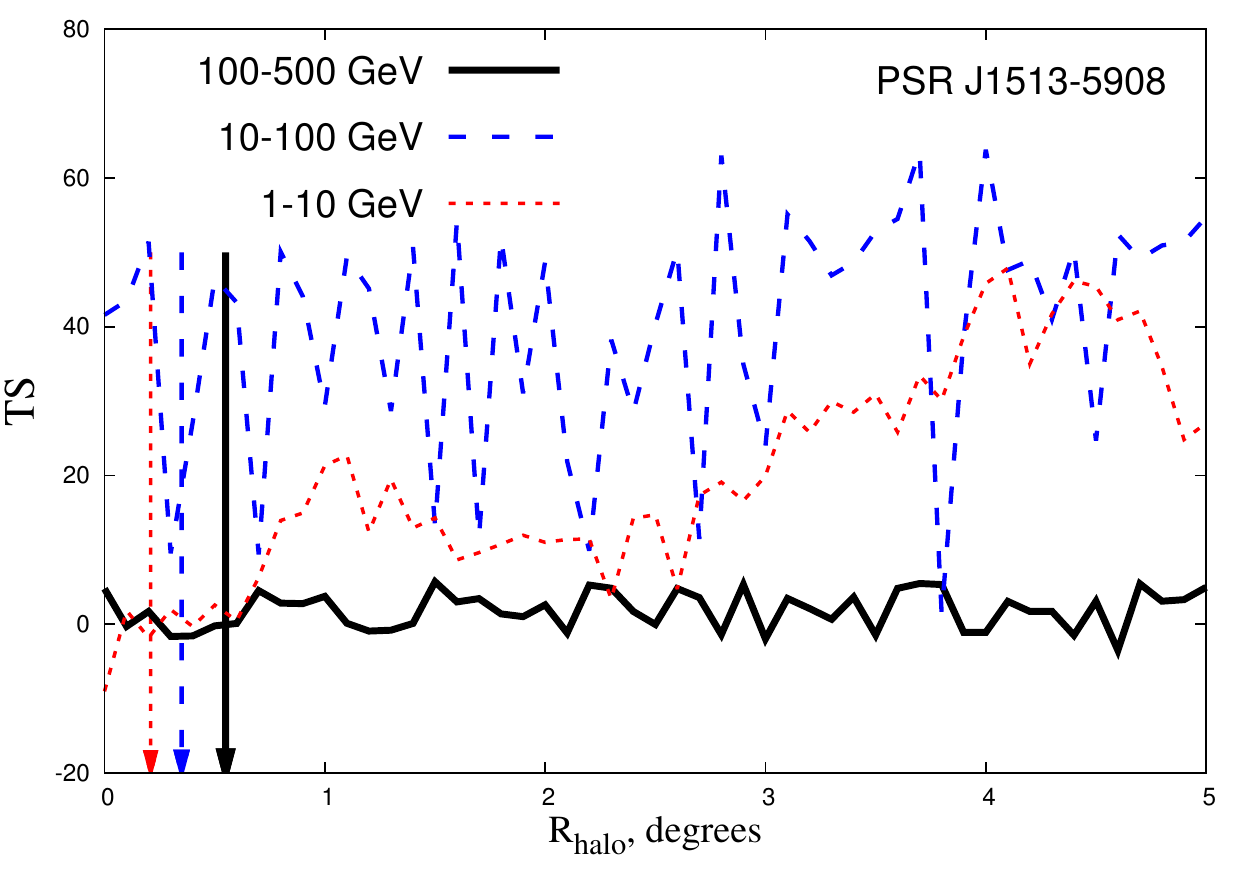}
\caption{ 
\label{fig5}
$TS(R_{halo})$ curves for Crab (PSR J0534+2200, upper panel) and 
PSR J1513-5908 (lower panel). 
Vertical arrows show the sizes of the halos 
that are expected from the estimate \eqref{eq:size1}.}
\end{figure}

\textbf{7)} 
The Crab Pulsar and Nebula are very bright gamma-ray sources in the Galaxy \cite{Aharonian:2006pe,Abdo:2009ec}. 
The Crab pulsar is also the youngest and the most energetic one from our sample, 
which is why it is most likely to feature a gamma-ray halo.
However, contrary to expectations, 
the analysis of the Crab pulsar does 
not show any evidence of a gamma-ray halo in any
energy bin (see the upper panel of Fig. \ref{fig5}). 
In the bins 10-100 and 100-500 GeV the $TS$ curves essentially coincide with the 
$TS=0$ axis. The $TS$ curve at 1-10 GeV does not show any smooth peak and oscillates around a constant value $TS\sim 40$.

\textbf{8)} 
The observation of the pulsar PSR J1513-5908 
(also known as PSR B1509-58)
did not disclose
any excess above 100 GeV (see the lower panel of Fig. \ref{fig5}).
The $TS$ curve in the range 10-100 GeV 
is very spiky and oscillates around a constant value $TS\simeq 30$
over the whole range of radii, which suggests that this offset is 
caused by uncertainties in background modeling. 
In the range 1-10 GeV
the TS curve has two peak-like features at $R_{halo}\approx 1^\circ$ and 
$R_{halo}\approx 4.5^\circ$. 
These peaks are far from being smooth and 
most likely are caused by other sources in the Galactic plane.
The pulsar of interest has rather low 
galactic latitude and adjoins many gamma-ray sources.
E.g., the $5^\circ$ region of interest contains 
at least four extended HESS sources of sizes $R\sim 0.1^\circ-0.3^\circ$ 
which were modeled 
as point sources in 3FGL 
(HESS J1503-582, HESS J1458-608, HESS J1458-608, HESS J1507-622), and a collective effect can mimic an extended halo.

\newpage

\vspace{0.4 cm}
\begin{center}
\textbf{B.1. Constraints on halo luminosity}
\end{center}
\vspace{0.2 cm}

Let us discuss now the constraints on halo luminosities, which can be obtained
from the subset of pulsars with compact halos. 
The most conservative bound can be derived by making use of
the pulsar PSR J1513-5908.
The scaling of the $TS$ with the halo flux [Eq.\eqref{eq:tscta_m}] suggests that 
the nonobservation of a halo in the energy bin 10-100 GeV at significance 
$TS=60$ can be translated into
the upper limit on the halo flux,
\be 
TS_{10-100}<60 \quad \Rightarrow \quad  F^{10-100\;\mathrm{GeV}}<  4\times 10^{-10} 
\;\mathrm{ph}/\mathrm{cm}^2\mathrm{s} \,.
\ee
Using the spectral index
$\Gamma=2.4$, this gives the constraints on the 
total flux and luminosity in the energy range $E_{\gamma}\geq 1$ GeV,
 \be
 \label{eq:const1513}
 \begin{split}
 &F^{E_\g \geq 1\;\mathrm{GeV}}<5.8\times 10^{-11}\;\mathrm{erg}/\mathrm{cm}^2\mathrm{s}\\
 &L_\g^{E_\g \geq 1\;\mathrm{GeV}}<1.3 \times 10^{35} {\mathrm{erg}}/{\mathrm{s}}\,,
 \end{split} 
 \ee
 which is 1 order of magnitude weaker than the bounds obtained for
 pulsars with large halos. 
 
 On the other hand, the strongest bound can be inferred from the Crab pulsar. Indeed,
 performing the same manipulations as above,
 one obtains
 \be 
 \label{eq:constCrab}
TS_{10-100}<10 \quad \Rightarrow \quad  L^{E_\g\geq 1\;\mathrm{GeV}}< 6.7\times 10^{33} 
\;\mathrm{erg}/\mathrm{s} \,.
\ee
The bounds on the luminosity for the rest of the pulsars with compact halos are scattered between those for 
Crab and PSR J1513-5908.

Notice that even the weakest bound \eqref{eq:const1513} 
is roughly an order of magnitude smaller
than the average luminosity of extended sources found 
in Ref.~\cite{Neronov:2012kz}. We will come back to this issue shortly.

\section{Discussion}
\label{sec:disc}

We found 
that only one pulsar (out of eight in our set)
has extended gamma-ray emission which may be 
interpreted as a CR halo. 
Yet this interpretation is far from being definitive,
which is why we stick to the constraints on halo luminosities
obtained for other pulsars and relate them to
the total energy of injected CRs.

We first focus on the pulsars with large halos. 
The constraints on the halo luminosity [cf. Eqs.~\eqref{eq:const0501},\eqref{eq:const1709},\eqref{eq:const2229}, and \eqref{eq:bin2large}]
are very similar for all of them, 
and can be written as
\be 
\label{eq:conlum}
L_\g^{halo}\lesssim (1-2)\times 10^{34} \;\mathrm{erg/s}\,.
\ee
The uncertainties induced 
by spectra extrapolations and inaccuracies in the scaling 
\eqref{eq:tscta_m} contribute only $\sim 30\%$ to the scatter
in Eq.~\eqref{eq:conlum}.

Using Eq.\eqref{eq:halolum},
the constraint \eqref{eq:conlum} can be translated into a constraint on the total cosmic-ray energy,
\be
\label{eq:const1}
\mathcal{E}^{halo}_{CR}\lesssim (0.5-1)\times 10^{50} \;\mathrm{erg}\,.
\ee
The above constraint implies that the total CR energy released by pulsars 
is still smaller than the benchmark mean value $\sim 2\times 10^{50}$ erg required in order to
produce the bulk of galactic CRs, although it hits the lower bound of Eq.\eqref{eq:pulsarate}.
Thus, at face value, our results disfavor the scenario in which all
galactic cosmic rays are injected in the ISM exclusively by newborn pulsars.
We note that our result accords with recent studies which imply that
the birth-period distribution of pulsars is close to log-normal with
a mean value $\sim$100 ms
\cite{Popov:2012ng,Noutsos:2013ce}.
This suggests that the rotational energy budget accessible for CR production
should be typically smaller than $10^{50}$ erg [see Eq.~\eqref{eq:erot}].

Before moving on we would like to comment more 
on the uncertainties in Eq.~\eqref{eq:const1}. 
First, one might worry about the diffusion coefficient $D$ in Eq.~\eqref{eq:diffcoeff},
which is uncertain by a factor of 3 due to the degeneracy with the height of the 
galactic CR halo \cite{Blasi:2011fi}. This coefficient enters
the size of the diffusive halo with a square root in 
Eqs.~\eqref{eq:size2} and \eqref{eq:size1}, so that the uncertainly in its value 
is only $\sqrt{3}\simeq 1.7$. 
This uncertainly can affect our splitting into large and compact halos
adopted in Sec.~\ref{sec:sample} by making 
the halos around the pulsars No 3 and 4 of Tables~\ref{tab:1} and~\ref{tab:halos} 
compact, which
calls for a reassessment of our constraints for these pulsars.
If these pulsars are indeed the ones with compact halos
one still can use 
the results in the energy bin 10-100 GeV Eq.\eqref{eq:bin2large},
which are taken into account in the constraint Eq.\eqref{eq:const1}.
To sum up, the uncertainty related to the diffusion constant appears to not be crucial for our analysis.

The second, and more serious source of degeneracy 
is the interstellar matter density, which explicitly affects the constraint \eqref{eq:const1} through Eq.\eqref{eq:halolum}.
Unfortunately, the measurements of density in the vicinity of pulsars are quite uncertain.
For the pulsars of interest the average ISM density lies in the range $0.3-1$~cm$^{-3}$ \cite{Kalberla},
which translates to the following scatter:
\be
\mathcal{E}^{halo}_{CR}\lesssim (0.5-3)\times 10^{50} \;\mathrm{erg}\,.
\ee
Our constraint now has an overlap with the energy required to 
account for all CRs exclusively with pulsars Eq.\eqref{eq:pulsarate}. We point out, however,
that the lower bound in Eq.\eqref{eq:pulsarate} is a conservative value which should be taken with
a grain of salt since it corresponds to a very high pulsar birthrate of $1/30 \;\mathrm{yr}^{-1}$. 
To sum up, the degeneracy between $\mathcal{E}^{halo}_{CR}$ and $n_{ISM}$ 
does not allow us to definitely rule out pulsars as main sources of CRs,
but our analysis indicates appreciable tension in this scenario.

As for the extended halo observed around PSR J0007+7303,
one might, in principle, interpret this gamma-ray emission 
as a counterpart of a CR halo.
In that case, comparing the luminosity of this halo \eqref{eq:lum007}
with Eq.~\eqref{eq:halolum} and using the density $n_{ISM}=~(0.05-0.1)~\text{cm}^{-3}$
\cite{Martin:2016uzs},
one can estimate the related energy budget of CRs,
\be
\label{eq:cons0}
\mathcal{E}_{CR}^{halo}\sim  (2-4)\times 10^{50} \; \mathrm{erg}.
\ee

Two comments are in order here. 
First, the interpretation of extended emission around PSR J0007+7303
as a SNR or PWN is not ruled out at the moment. 
Thus, the value given in Eq.\eqref{eq:cons0} should be considered as a conservative upper bound,
since by having accounted for the presence of a SNR and PWN one will inevitably get a stronger constraint.
Second, the emission in the energy bin 1-10 GeV
can also be produced by electrons or positrons via inverse Compton scattering.
The halo's angular radius projected at the pulsar's distance yields the physical 
halo size $\sim 30$ pc, which implies that the lepton contribution can be quite 
significant at such small distances from the source.
In order to clarify the situation an additional multiwavelength analysis
of this halo is needed.

Now we discuss the subpopulation of pulsars with compact halos, 
which are younger and more energetic than the four we discussed above.
Since the halo sizes are expected to be quite small for these pulsars,
their observation with LAT becomes challenging given its limited resolution at small angular scales.
That is why we expect the constraints on CR power to degrade
if the analysis is based only on the youngest pulsars. 
The bounds on the halo luminosity \eqref{eq:const1513}-\eqref{eq:constCrab} 
can be related to 
the total energy of CRs via Eq.~\eqref{eq:halolum},
\be
\mathcal{E}_{CR}^{halo}\lesssim (0.3 - 7)\times 10^{50}\;\mathrm{erg}\,.
\ee
The range is quite wide in this case and,
if we used only the subpopulation of pulsars with compact halos,
the scenario in which all CRs in the Galaxy are born by pulsars would be largely
unconstrained.

One might notice the apparent tension between our results and 
the detection of gamma-ray halos above 100 GeV 
with average fluxes $\sim 5\times 10^{-11}$ erg/cm$^2$s reported 
in Ref.~\cite{Neronov:2012kz}.
These fluxes yield the typical halo luminosity above 1 GeV $\sim 5\times 10^{35}$ erg/s [see Eq.\eqref{eq:nslum}],
significantly exceeding our bounds.
There are several ways to explain this tension.

On the one hand, the extended halos observed in Ref.~\cite{Neronov:2012kz} may
be spurious, i.e., produced
by background fluctuations or projection effects.
With the new Fermi-LAT data we checked that the {\it N-S} 
emission is not due to background fluctuations
(see Appendix~\ref{app:fluct} for details).
The interpretation of {\it N-S} halos
as a projection effect of several independent VHE sources,
however, cannot be excluded, and moreover seems plausible 
given that all the {\it N-S} sources are located in the Galactic plain.

On the other hand, the halos observed in Ref.~\cite{Neronov:2012kz}
can be produced by a mechanism involving 
multiple sources, such as the interaction between pulsars, SNRs,
and the interstellar medium. In such a case 
these halos can exist only in a specific environment and 
there is little hope to find them in each sample of young pulsars. 
This explanation is supported by the fact that most of the {\it N-S}
sources were found in the Norma arm of the Galaxy, 
which is known as a peculiar region with the 
highest star-formation rate and an average gas density $n_{ISM}\sim 10$~cm$^{-3}$~\cite{Bronfman}.
The high density of ISM and the presence of molecular clouds
can significantly boost the luminosity of extended halos 
even with the CR input satisfying Eq.~\eqref{eq:const1},
which can readily resolve the tension. It should be stressed that ``boosted luminosity" does not mean an increase in total energy, i.e., pulsars still cannot be the main source of CRs.
More studies of the {\it N-S} sources 
in other frequency bands are needed
in order to further clarify the situation.

Our analysis disfavors the pulsar origin of the bulk of galactic CRs,
but it does not pin down the scenario in which 
pulsars produce only very energetic CRs with $E_{CR}\gtrsim 100$ GeV,
while other mechanisms are responsible for particle acceleration at
lower energies.
This scenario may well be true in the light of new evidence that
the interstellar CR spectrum has multiple 
components~\cite{Neronov:2011wi,Adriani:2011cu}.
This scenario can also reconcile the mentioned tension with Ref.~\cite{Neronov:2012kz}
since our constraints are, essentially,
based on the events in the energy range 1-10 GeV,
which should be mostly due to CR with $E_{CR}\lesssim 100$ GeV,
though in this case the spectral shape of CRs produced in pulsars should be rather specific.

Our analysis may be extended in several ways. One can further investigate
the extended emission that we observed around 
PSR J0007+7303/SNR CTA1, and test it for signatures of CR production.
Another way to go is to study in detail 
the nature of the extended emission found in Ref.~\cite{Neronov:2012kz}.
As discussed, this requires proper accounting for VHE sources in their vicinity,
and the use of other energy bands and neutrino signals \cite{Neronov:2013lza}. 
Also it will be interesting to update 
our analysis once more data are accumulated, e.g., 
with the new gamma-ray telescopes such as CTA or HAWC
\cite{Acharya:2013sxa,DeYoung:2012mj}.

\section{Conclusions}
\label{sec:concl} 
 
In this paper we scrutinized the hypothesis that 
galactic CRs are produced by pulsars at birth.
In order to account for the bulk of the galactic CRs it is sufficient 
that their sources release some $\sim  2\times 10^{50}$ erg
energy every $\sim 50$ years in the form of CRs. This power can be, in principle,
generated by the rotational energy of neutron stars right after 
supernova explosions.
If this is the case, CRs should interact with the
interstellar medium as they escape from their parent pulsars and thus produce 
gamma radiation observable as extended halos. 
The observations of these halos can be used in order to estimate 
the total energy of injected CRs.

In this study we sought gamma-ray halos around young 
pulsars in the recent 
7-year Fermi-LAT data.
Using the Pass 8 reconstruction and statistical tools provided by the LAT Collaboration,
we tested a specially selected 
sample of pulsars whose 
hypothetical gamma-ray halos could be unambiguously identified.
As a result, we found only one extended source which can be interpreted
as a gamma-ray counterpart of a CR halo.
This is the one-degree halo around 
the pulsar PSR J0007+7303 detected in the energy bin 1-10 GeV.
The overall
luminosity of the halo above 1 GeV is
$\sim 3\times 10^{33}$ erg/s,
which implies the total energy of corresponding CRs
$\sim 2\times 10^{50}$ erg.
We emphasize
that the other interpretations of this emission are not excluded
and further studies of this source are required.

Without any assumptions on the nature of this emission
we derived a constraint on the typical luminosities 
of gamma-ray halos, $L_{halo}\lesssim 10^{34}$ erg/s. 
This implies that the total energy of CRs 
produced by a pulsar at birth should typically be smaller than
$10^{50}$ erg, and thus, disfavors the scenario in which 
galactic CRs are produced entirely by pulsars. 
There are possible caveats in the interpretation of our result. 
First, our constraints are quite degenerate with the ISM density. 
Second, there is large uncertainty
in the expected pulsar CR luminosity due to current imperfect knowledge of pulsar birthrates.

\section*{Acknowledgements} 

We thank M.~Libanov, S.~Sibiryakov, B.~Nizamov, S.~Troitsky and V.~Vasiliev for useful discussions.
We thank K.~Postnov for pointing out the references on the initial periods of pulsars, 
and for helpful and encouraging comments.
We are grateful to A.~Neronov and D.~Semikoz for their valuable comments on the draft.
We also thank the referee for a thorough 
report which has substantially improved the presentation.
The authors acknowledge the
support by the Russian Science Foundation grant 14-12-01340.
The analysis is based on the data and software provided by the Fermi Science Support Center 
(FSSC).
The numerical part of the work has been performed at the cluster of the Theoretical Division of INR RAS.
During the work on this paper the authors were using 
the SIMBAD database
and the ATNF pulsar database.

\appendix

\section{N-S pulsars and their  properties}
\label{app:NSpuls}

\begin{table}[h]
\begin{tabular}{|c|c|c|c|c|c|c|}
\hline
PSR  & $l$ & $b$ & $r_s$, kpc & $T_{SD}$, kyr & $\dot E$, erg/s &$P$, s \\\hline
 B1800-21 & $8.40 $ & $0.15$ & $4.40$ & $15.8$&$2.2\times 10^{36}$ &$0.13$\\\hline
 B1823-13 & $18.00  $ & $-0.69$ & $4.12$ & $21.4$& $2.8\times 10^{36}$&$0.10$\\\hline
 J1838-0655 & $25.25 $ & $-0.20$ & $6.60$ & $22.7$& $5.5\times 10^{36}$&$0.07$\\\hline
 J1841-0524 & $27.02  $ & $-0.33$ & $4.89$ & $30.2$& $1.0\times 10^{35}$&$0.44$\\\hline
 J1856+0245 & $36.01 $ & $0.06$ & $10.29$ & $20.6$& $4.6\times 10^{36}$&$0.08$\\\hline
 J2021+4026 & $78.23 $ & $2.09$ & $2.15$ & $76.9$& $1.2\times 10^{35}$&$0.26$\\\hline
 J1023-5746& $284.17$ & $-0.41$ & $ * $ & $4.6$& $1.1\times 10^{37}$&$0.11$\\\hline
J1420-6048& $313.54$ & $0.23$ & $7.65$ & $13$& $1.0\times 10^{37}$&$0.068$\\\hline
 J1614-5144& $331.62$ & $-0.58$ & $9.56$ & $3270$& $8.1\times 10^{31}$&$1.5$\\\hline
 J1617-5055& $332.50$ & $-0.28$ & $ 6.46$ & $8.13$& $1.6\times 10^{37}$&$0.069$\\\hline
 J1632-4757& $336.30$ & $0.08$ & $ 6.96$ & $24$& $5.0\times 10^{34}$&$0.23$\\\hline
 J1648-4611& $339.44$ & $-0.79$ & $ 5.71$ & $110$& $2.1\times 10^{35}$&$0.16$\\\hline
 J1702-4128 & $344.74$ & $0.12$ & $ 5.18$ & $55.1$& $3.4\times 10^{35}$&$0.18$\\\hline
 J1708-4008 & $346.48$ & $0.04$ & $ 3.80$ & $8.9$& $5.8\times 10^{32}$&$11.0$\\\hline
 B1830-08 & $23.39$ & $0.06$ & $ 4.50$ & $147$& $5.8\times 10^{35}$&$0.085$\\\hline
\end{tabular}
\caption{The pulsars coincident with the extended sources in Ref.~\cite{Neronov:2012kz}.}
\label{tab:ns}
\end{table}

The {\it N-S} pulsars (those listed in Table II of Ref.~\cite{Neronov:2012kz}) and their characteristics 
are displayed 
in Table~\ref{tab:ns}.\footnote{Notice that the pulsar PSR J1708-4008 is incorrectly written  
in Table II of Ref.~\cite{Neronov:2012kz} as PSR J1706-4009.}
In order to check that our set of pulsars belongs to the same population as the {\it N-S} pulsars,
we first imposed a cut $T_{SD}<30$ kyr which selected 9 pulsars out of 15 present in Table~\ref{tab:ns}.
Then we applied the two-sample KS test for a selected set of {\it N-S} pulsars and our set (see Table~\ref{tab:1}).
We performed this test for the distributions over $\dot E$ and $P$ separately, 
and found the following p-values for either case:
\be
p_{KS}(\dot E)=0.84\,,\quad  p_{KS}(P)=0.62\,.
\ee

\section{Simulations of gamma-ray halos}
\label{app:simul}

In this appendix we discuss in detail the simulations performed in order to better understand the 
potential signal. We chose to simulate 
the pulsar No 1 (PSR J0007+7303) for pulsars with large halos and
the pulsar No 8 (PSR J1513-5908) for pulsars with compact halos.

\begin{table}[h!]
\begin{tabular}{|c|c|c|c|c|}
\hline
E, GeV & $R_{halo}$  & $F$,  cm$^{-2}$s$^{-1}$ [$\Gamma=2.4$] &  $F$, cm$^{-2}$s$^{-1}$ [$\Gamma=2$] \\\hline
100 - 500 & $0.6^\circ $ & $2.2\times 10^{-10}$ &  $2.0\times 10^{-10}$\\\hline
10 - 100 & $0.4^\circ $ & $5.8\times 10^{-9}$ &$2.2\times 10^{-9}$ \\\hline
1 - 10 & $0.2^\circ $ & $1.5\times 10^{-7}$ & $2.2\times 10^{-8}$\\\hline
\end{tabular}
\caption{Fluxes (in photons/cm$^2$/s) and angular sizes of the simulated {\it bright} gamma-ray halo around the pulsar PSR J1513-5908 for different energy bands and spectral indices.}
\label{tab:bright1513}
\end{table}
\begin{table}[h!]
\begin{tabular}{|c|c|c|c|c|}
\hline
E, GeV & $R_{halo}$  & $F$,  cm$^{-2}$s$^{-1}$ [$\Gamma=2.4$] &  $F$, cm$^{-2}$s$^{-1}$ [$\Gamma=2$] \\\hline
100 - 500 & $5.0^\circ $ & $5.0\times 10^{-10}$ &  $5.0\times 10^{-10}$\\\hline
10 - 100 & $3.2^\circ $ & $1.4\times 10^{-8}$ &$5.6\times 10^{-9}$ \\\hline
1 - 10 & $2.0^\circ $ & $3.4\times 10^{-7}$ & $5.6\times 10^{-8}$\\\hline
\end{tabular}
\caption{Fluxes (in photons/cm$^2$/s) and angular sizes of the simulated {\it bright} gamma-ray halo around the pulsar PSR J0007+7303 for different energy bands and spectral indices.}
\label{tab:bright0007}
\end{table}
\begin{table}[h!]
\begin{tabular}{|c|c|c|c|c|}
\hline
E, GeV & $R_{halo}$  & $F$,  cm$^{-2}$s$^{-1}$ [$\Gamma=2.4$] &  $F$, cm$^{-2}$s$^{-1}$ [$\Gamma=2$] \\\hline
100 - 500 & $0.6^\circ $ & $3.7\times 10^{-11}$ &$1.0\times 10^{-10}$\\\hline
10 - 100 & $0.4^\circ $ & $1.0\times 10^{-9}$ &$1.0\times 10^{-9}$ \\\hline
1 - 10 & $0.2^\circ $ & $2.5\times 10^{-8}$ &$1.0\times 10^{-8}$\\\hline
\end{tabular}
\caption{Fluxes (in photons/cm$^2$/s) and angular sizes of the simulated {\it faint} gamma-ray halo around the pulsar PSR J1513-5908 for different energy bands and spectral indices.}
\label{tab:faint1513}
\end{table}
\begin{table}[h!]
\begin{tabular}{|c|c|c|c|c|}
\hline
E, GeV & $R_{halo}$  & $F$,  cm$^{-2}$s$^{-1}$ [$\Gamma=2.4$] &  $F$, cm$^{-2}$s$^{-1}$ [$\Gamma=2$] \\\hline
100 - 500 & $5.0^\circ $ & $7.4\times 10^{-12}$ &  $4.44\times 10^{-11}$\\\hline
10 - 100 & $3.2^\circ $ & $2.0\times 10^{-10}$ &$5.0\times 10^{-10}$ \\\hline
1 - 10 & $2.0^\circ $ & $5.0\times 10^{-9}$ & $5.0\times 10^{-9}$\\\hline
\end{tabular}
\caption{Fluxes (in photons/cm$^2$/s) and angular sizes of the simulated {\it faint} gamma-ray halo around the pulsar PSR J0007+7303 for different energy bands and spectral indices.}
\label{tab:faint0007}
\end{table}

The pulsar PSR J0007+7303 is very close and its galactic latitude is rather high
($b\sim 10^\circ$), which results in a very low density of the LAT gamma-ray sources in the $10^\circ$ RoI 
around this pulsar (there are only 22 sources).
This pulsar thus represents the most clear case for our study.
As discussed above, the hypothetical halo around this pulsar should have very large angular extension; see Table~\ref{tab:halos}.

\begin{figure}[h!]
\includegraphics[width=0.5\textwidth]{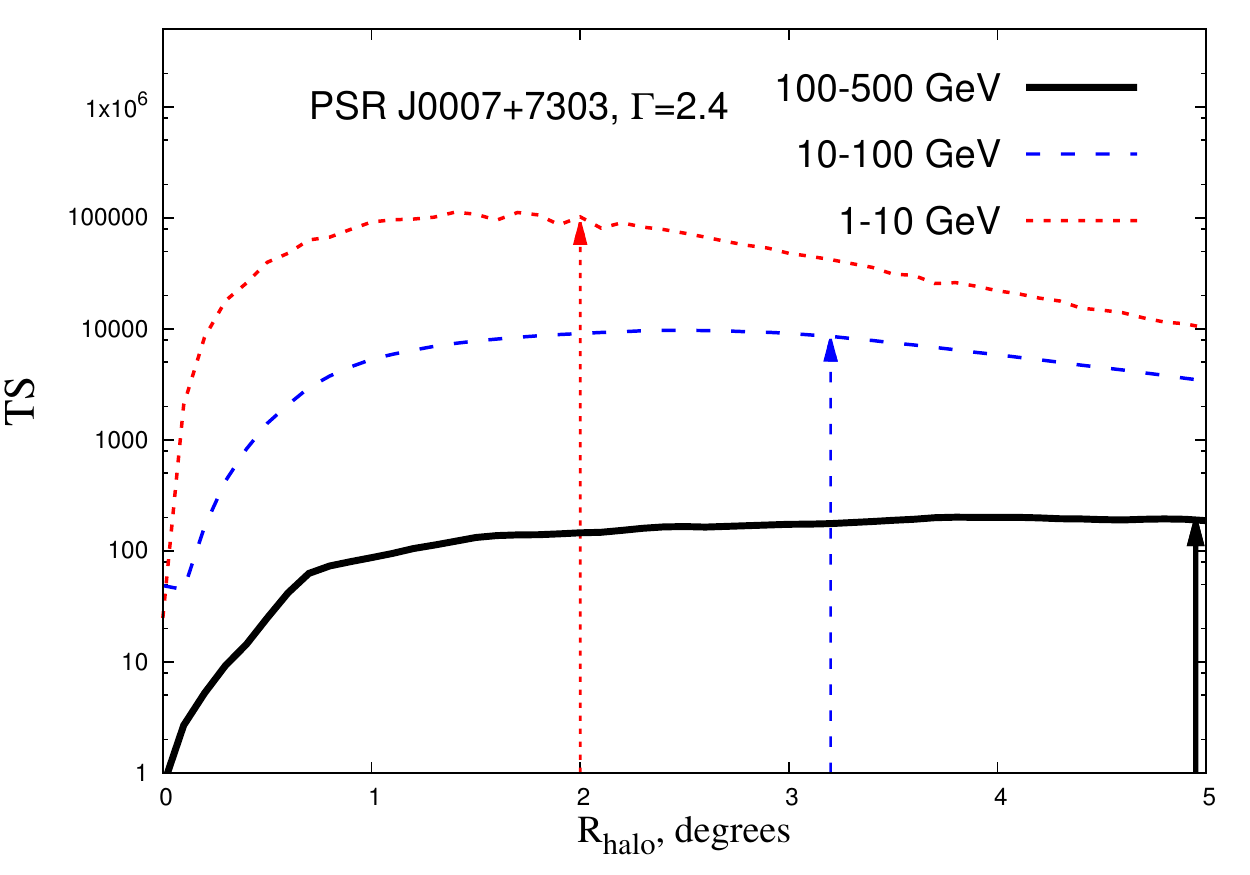}
\includegraphics[width=0.5\textwidth]{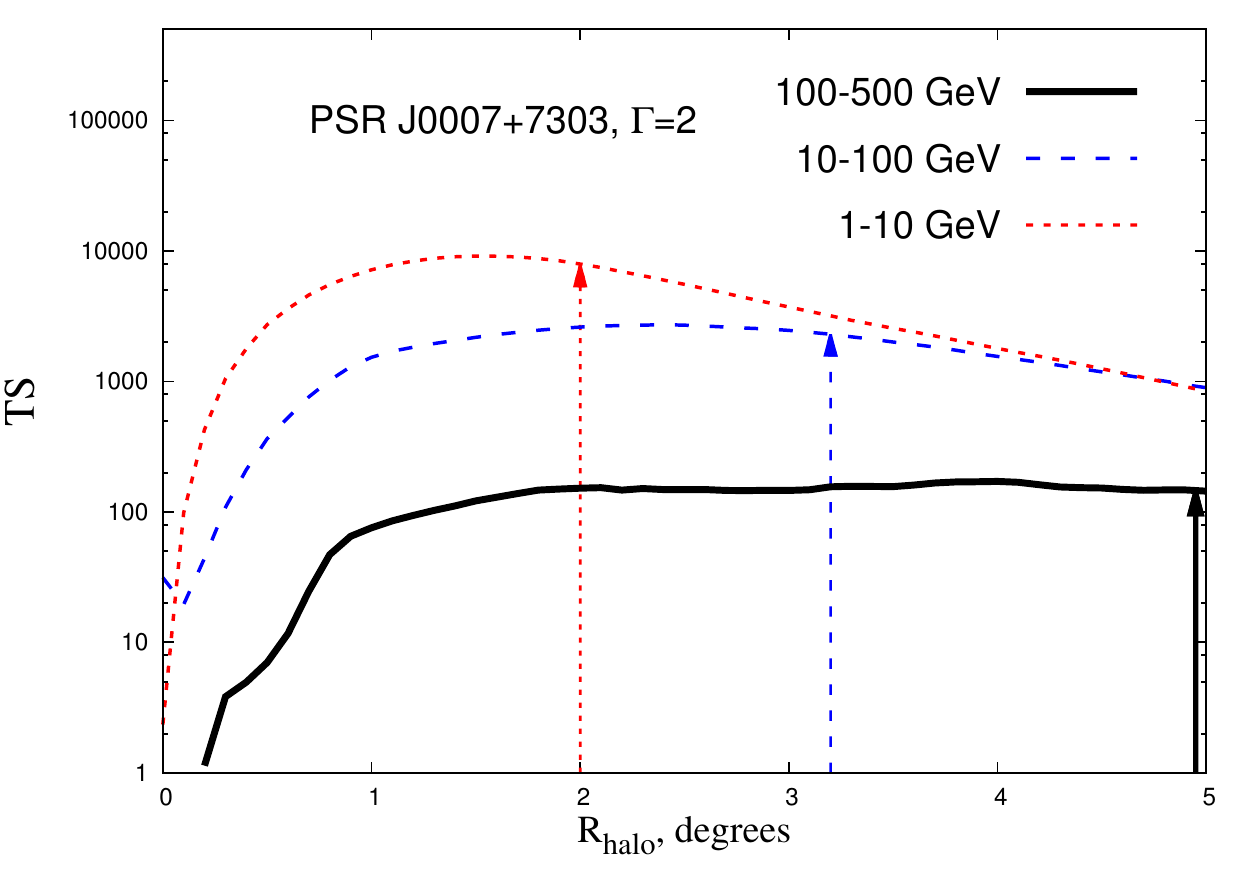}
\caption{\label{fig:sim_007_bright}
$TS(R_{halo})$ curves for 
the simulated {\it bright} gamma-ray halo around the pulsar PSR J0007+7303
for the spectral indices $\Gamma=2.4$ (upper panel) and
$\Gamma=2$ (lower panel).
The results of the analysis in different energy bands
are shown as a black solid line for 100-500 GeV,  
a blue dashed line for 10-100 GeV, and a red dotted line for 1-10 GeV. 
Vertical arrows show the sizes of the halos 
that were input 
into the simulations.
}
\end{figure}

\begin{figure}[h!]
\includegraphics[width=0.5\textwidth]{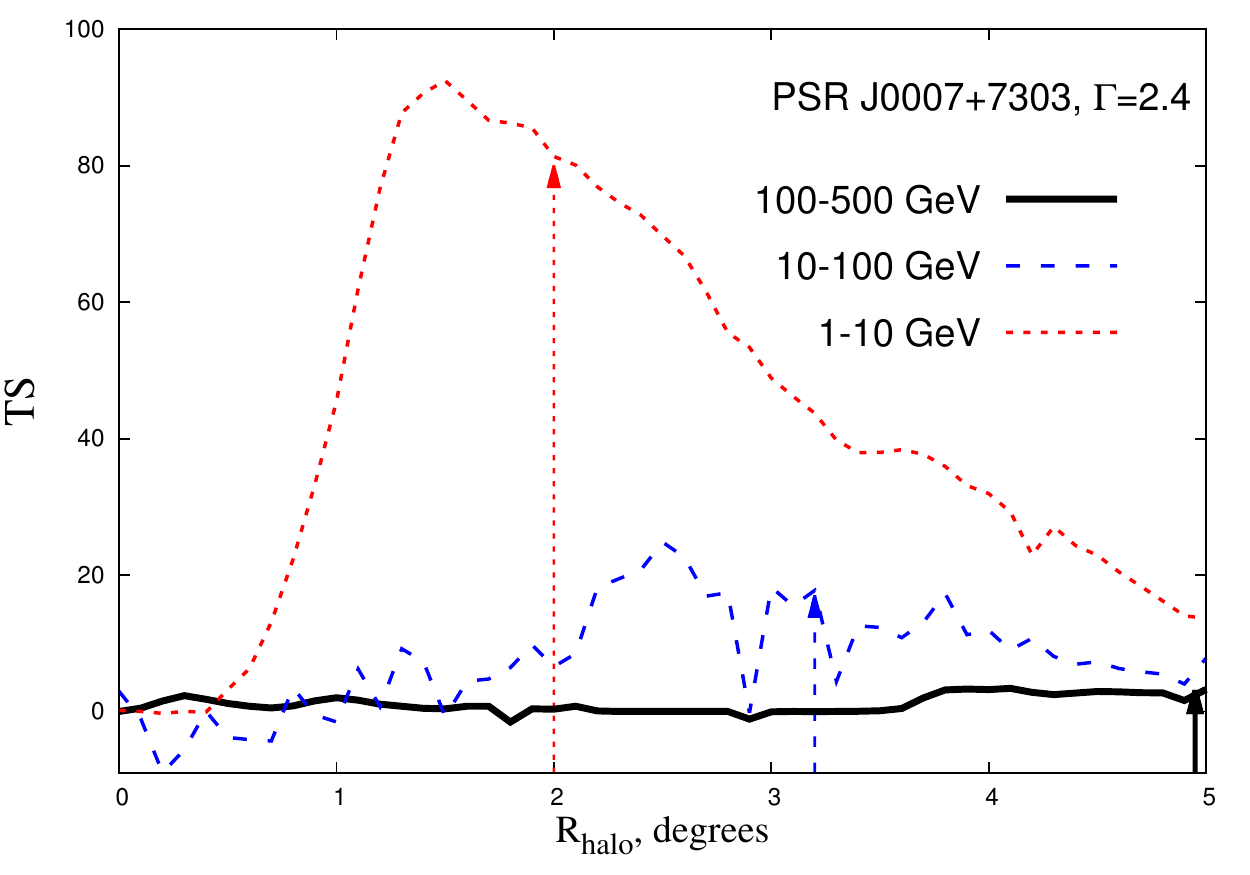}
\includegraphics[width=0.5\textwidth]{cta_red_3003G2}
\caption{\label{fig:sim_007_faint}
$TS(R_{halo})$ curves for 
the simulated {\it faint} gamma-ray halo around the pulsar PSR J0007+7303
for the spectral indices $\Gamma=2.4$ (upper panel) and
$\Gamma=2$ (lower panel).
The results of the analysis in different energy bands
are shown as a black solid line for 100-500 GeV,  
a blue dashed line for 10-100 GeV, and a red dotted line for 1-10 GeV. 
Vertical arrows show the sizes of the halos 
that were inserted 
into the simulations.
}
\end{figure}

\begin{figure}[h]
\includegraphics[width=0.5\textwidth]{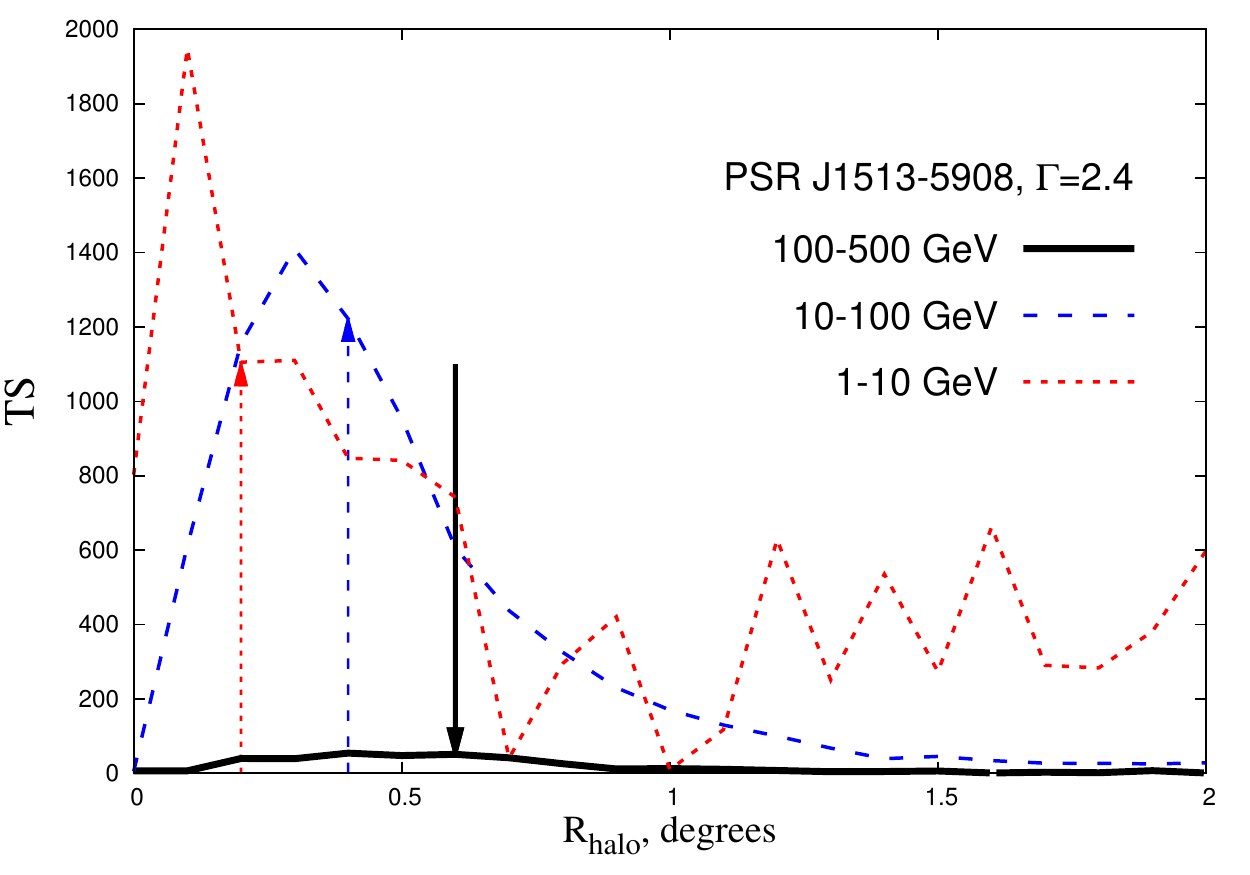}
\includegraphics[width=0.5\textwidth]{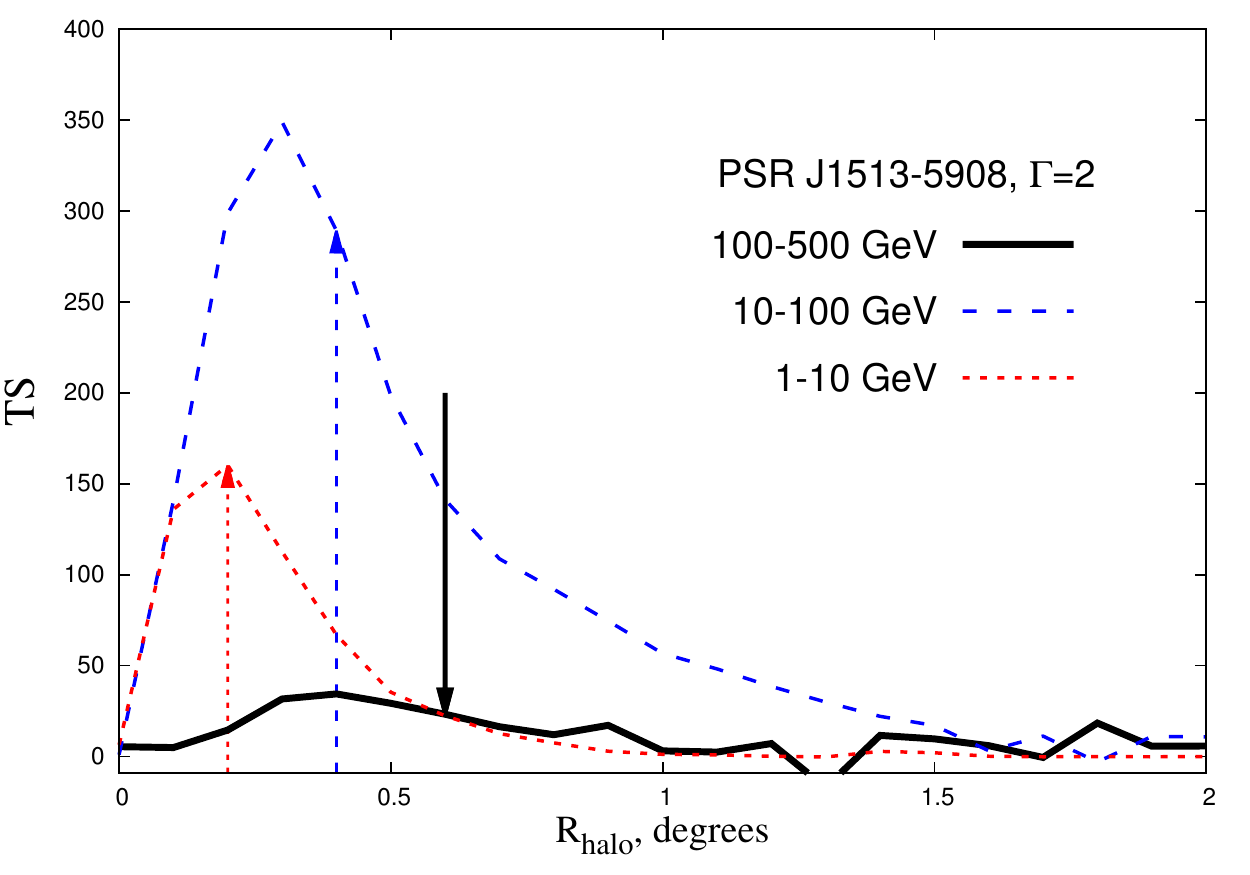}
\caption{\label{fig:sim_1513_bright}
$TS(R_{halo})$ curves for 
the simulated {\it bright} halo around the pulsar PSR J1513-5908
for the spectral indices $\Gamma=2.4$ (upper panel) and
$\Gamma=2$ (lower panel).
}
\end{figure}

\begin{figure}[h]
\includegraphics[width=0.5\textwidth]{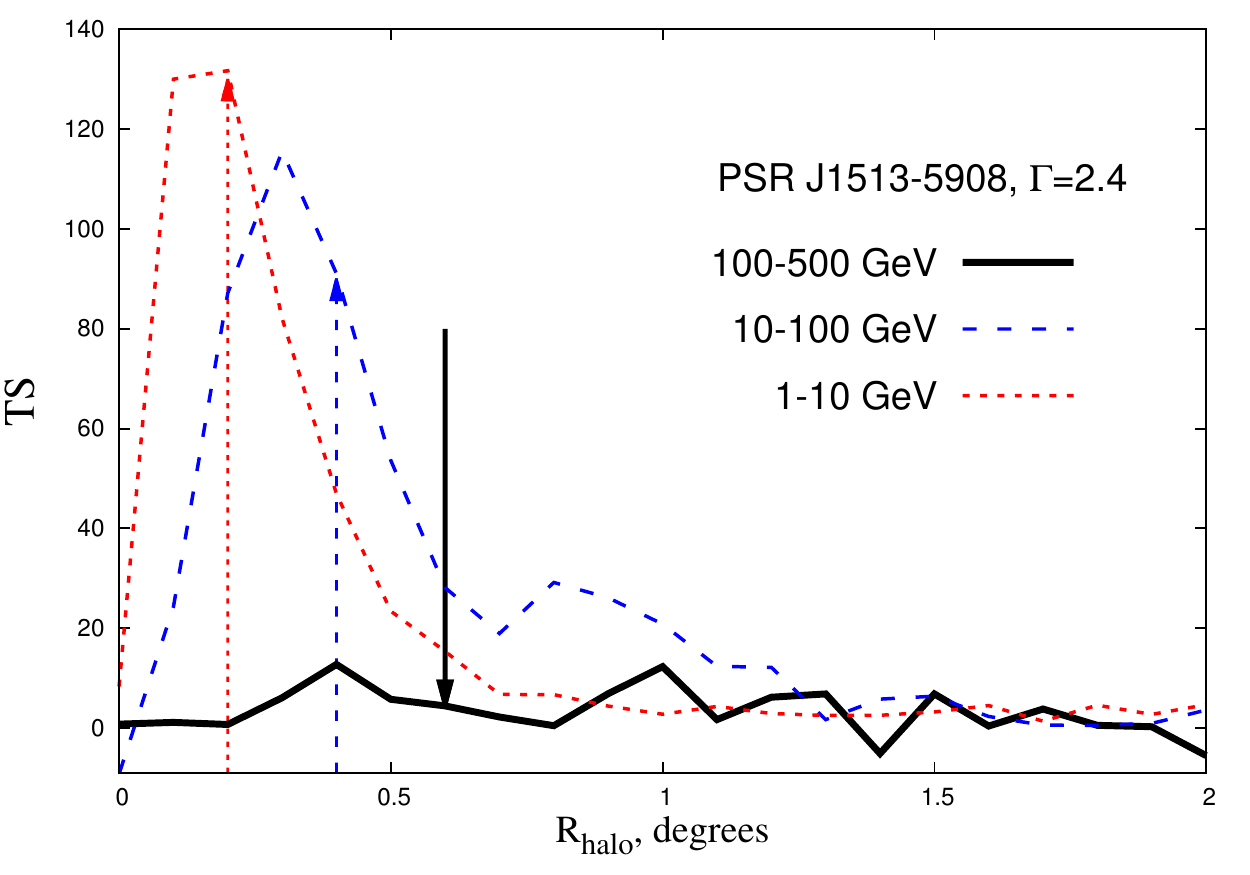}
\includegraphics[width=0.5\textwidth]{1513_red_704_G2}
\caption{\label{fig:sim_1513_faint}
$TS(R_{halo})$ curves for 
the simulated {\it faint} halo around the pulsar PSR J1513-5908
for the spectral indices $\Gamma=2.4$ (upper panel) and
$\Gamma=2$ (lower panel).
}
\end{figure}

On the other hand, the pulsar PSR J1513-5908 is pretty far away
and located 
close to the Galactic plane. 
Apart from the pulsar itself, 
there are 51 other LAT sources 
in the corresponding $10^\circ$ RoI.
The pulsar of interest is very young ($T_{SD}\sim 1$ kyr) and has a significant energy loss rate ($\dot E \sim 10^{37}$ erg/s).
The halo around this pulsar should be quite small; see Table~\ref{tab:halos}.

To generate the Fermi-LAT events we made use of 
the {\it gtobssim} utility. 
For either pulsar we simulated events in the energy range 1-500 GeV
for the relevant time interval (361 weeks)
in the 10$^\circ$ RoI around the pulsar.
The input model included all the LAT point and extended sources
located within the RoI, the galactic and isotropic background, and the gamma-ray halo
around the chosen pulsar.
Spectral parameters and photon fluxes for the 3FGL sources were taken
directly from the 3FGL catalogue, and the recommended values were chosen for the 
isotropic and galactic background fluxes.\footnote{http://fermi.gsfc.nasa.gov/ssc/data/analysis/scitools/help/ gtobssim.txt}

As discussed in Sec.~\ref{sec:simul}, for either pulsar we simulated two types of 
halos: the {\it bright} one and the {\it faint} one.
For the bright halos [case {\it (a)} in what follows] 
we assumed the fluxes of order \eqref{eq:flux} in the energy bin 100-500 GeV. 
Having fixed the flux in the range 100-500 GeV [Eq.~\eqref{eq:flux}] and assuming a simple power-law spectrum
of a halo, ${dN}/{dE}= A_0 \times E^{-\Gamma}$, we computed the normalization factor $A_0$ for two particular choices of 
the spectral index: $\Gamma=2.4$ and $\Gamma=2$. 
This yielded the halo fluxes in the energy 
bands 1-10 GeV and 10-100 GeV.

In the case of a bright halo around the pulsar PSR J1513-5908 we 
fixed the flux \eqref{eq:flux} at $r_s=4.4$ kpc in the energy bin 100-500 GeV
and extrapolated the spectrum down to 1 GeV as discussed above. 
The results are shown 
in the two right columns of Table~\ref{tab:bright1513}. 
For the pulsar PSR J0007+7303, in fact, 
the straightforward substitution $r_s=1.4$ kpc in Eq.~\eqref{eq:flux} yielded
a very high flux value. The halo appeared 
to be so bright that it drastically 
deteriorated
the convergence of our likelihood optimization procedure. 
In order to facilitate the numerical analysis for this pulsar,
we reduced the flux 4 times compared to the one extracted directly from 
Eq.~\eqref{eq:flux}. The resulting fluxes are listed in Table~\ref{tab:bright0007}.

In the case of faint halos [case {\it (b)} in what follows]
we were looking for typical fluxes that yield the detection at $TS\sim 100$
in one of the energy bins.

For PSR J1513-5908 we found that the flux $10^{-9}$ ph/cm$^2$s
in the energy bin 10-100 GeV gives the halo detection at $TS\sim 120$.
Having fixed the flux in this range, we derived the fluxes at 1-10 GeV and 100-500 GeV for spectra with $\Gamma=2.4$ and $\Gamma=2$.
The results are listed in the two right columns of Table~\ref{tab:faint1513}.

In the case of PSR J0007+7303 we found that the flux 
$5\times 10^{-9}$ ph/cm$^2$s in the energy bin 1-10 GeV
leads to halo detection at $TS\sim 100$ in this range.
Then, having
fixed the flux at 1-10 GeV and assuming a power-law
spectrum with indices $\Gamma=2.4$ and $\Gamma=2$, 
we computed the fluxes in the bins 10-100 GeV and 100-500 GeV.
The results are shown in the two right columns of Table~\ref{tab:faint0007}.

Given the power-law 
distribution of photons, most of them ``sit" at the 
lower boundary of each energy bin.
Thus, it is natural to assume that
the angular size of the halo within each narrow 
energy bin is constant and defined by the lower energy of the bin.\footnote{In fact, the photons with higher energy have bigger statistical significance. However, this subtlety is not crucial for our further analysis given other uncertainties in the estimate \eqref{eq:size1}.}
In that way, we computed the sizes of the gamma-ray halos in the energy bins 1-10, 
10-100, and 100-500 GeV by plugging 
the values $E_{\gamma}=1,10,100$ GeV, correspondingly, into Eq.~\eqref{eq:size1}.
The obtained angular sizes of the gamma-ray halos 
for relevant energy ranges are listed in the second columns of 
Tables~\ref{tab:bright0007},\ref{tab:bright1513},\ref{tab:faint0007},
and \ref{tab:faint1513}.

The gamma-ray halo was inserted into the {\it gtobssim} 
source models for both pulsars as three uniformly bright 
circles of sizes and fluxes given in 
Tables \ref{tab:bright0007},\ref{tab:bright1513},\ref{tab:faint0007},
and \ref{tab:faint1513},
such that the flux of each template was restricted to 
the corresponding energy band and was put to zero
everywhere else. 
Within each band the photons were distributed
over the power law with the corresponding index $\Gamma$.

After generation, the simulated events were processed using the 
{\it gtlike} utility analogous to the real data (see Sec.~\ref{sec:method}).

{\bf 1a)}
The results of the analysis for the simulated bright gamma-ray halo around PSR J0007+7303
are shown in Fig.~\eqref{fig:sim_007_bright}.
We see that the {\it gtlike} utility is more
biased towards smaller halo sizes than the simulated
ones, which indicates that the likelihood optimization procedure prefers 
halos with larger surface brightness.
The bias is quite strong above 100-500 GeV (where less events are present)
and weakens at lower energies which contain more statistics.
In the bin 100-500 GeV the $TS$ curve 
flattens already at $R_{halo}\simeq 2^\circ$ and turns into a plateau with $TS\sim 160$.
In the other energy bins (1-10 GeV and 10-100 GeV)
the $TS$ curves also become flat rather fast and after that 
have very moderate dependence on $R_{halo}$.

{\bf 1b)}
The behavior is similar in the case of the faint gamma-ray halos around PSR J0007+7303
(see Fig.~\ref{fig:sim_007_faint}), where one 
can still see a small offset between the sizes of 
simulated and observed halos. 
Note that even in the case of a faint halo, 
the detection with significance $TS\simeq 20$ ($TS\simeq 60$) is possible 
in the energy bin 10-100 GeV
for the power-law index $\Gamma=2.4$ ($\Gamma=2$). The significance in the range 100-500 GeV is very small, which means that a halo is practically undetectable in this bin.

{\bf 2a)}
The results of our analysis of the bright halo around PSR J1513-5908 for the spectral indices 
2 and 2.4 are displayed in Fig. \ref{fig:sim_1513_bright}.
The small size of the halo with respect to the LAT PSF 
results in a relatively 
small statistical significance of the halo at energies 1-10 GeV.
We found, again, a small bias between the simulated and detected sizes 
of the halo at energies above 10 GeV. 
This bias is, however, not as strong as the one we saw in the PSR J0007+7303 case.
Given that for either spectral index we fixed the same flux above 100 GeV, 
the signal is very similar in this energy range and has $TS\sim 40$ for either spectrum.
The excess in the energy bin 10-100 GeV is quite significant in either case.

The signal in the energy bin 1-10 GeV is quite different depending on the spectral index.
A soft spectrum with $\Gamma=2.4$ implies a larger signal in the energy bin 1-10 GeV,
which has the same $TS$ as the signal at 10-100 GeV
(see the upper panel of Fig.~\ref{fig:sim_1513_bright}).
On the other hand, a harder spectrum with 
$\Gamma=2$ implies a rather faint signal in the energy bin 1-10 GeV.

\begin{figure}[h]
\includegraphics[width=0.5\textwidth]{scal_all}
\includegraphics[width=0.5\textwidth]{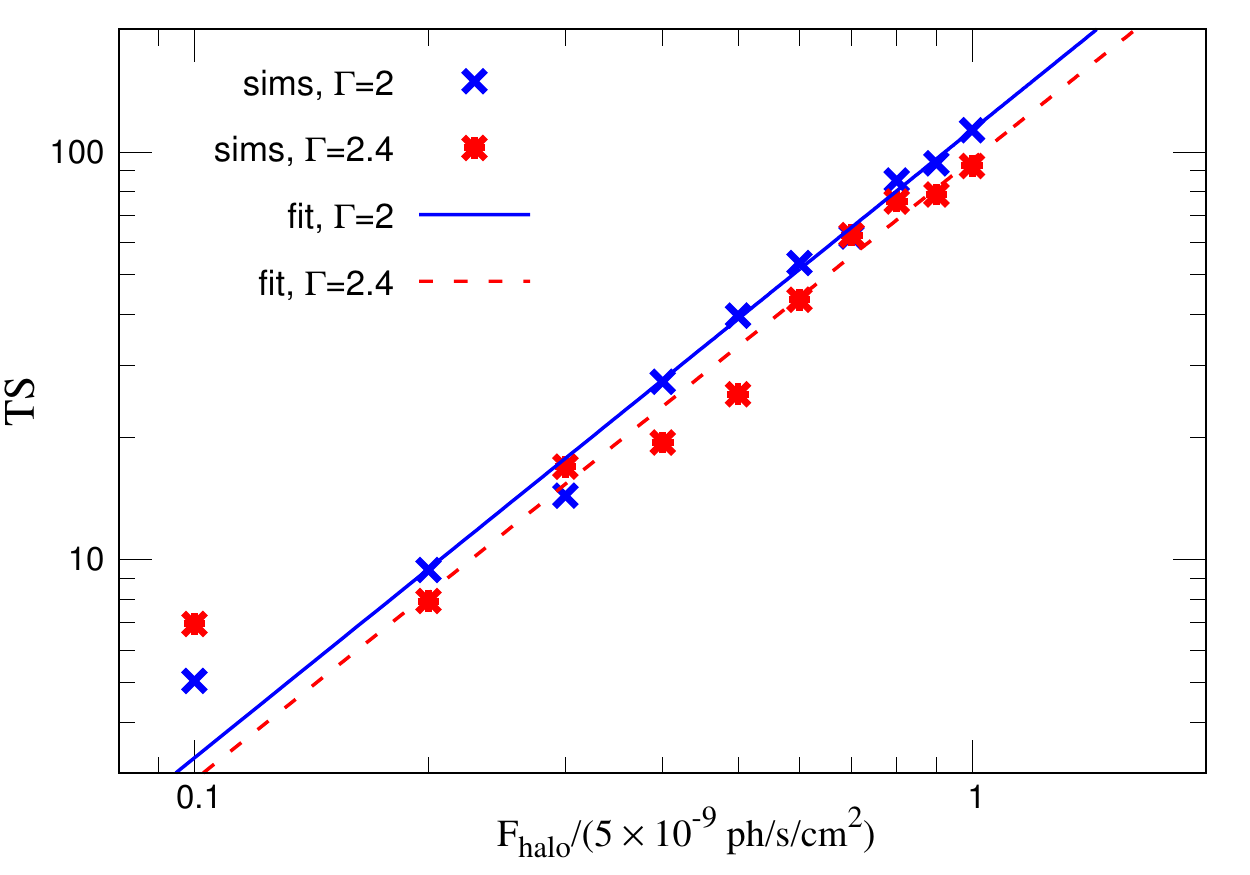}
\caption{\label{fig:ts_scal}
The $TS(F_{halo})$ dependence retrieved from simulations. 
Upper panel: Results for different energy bins for $\Gamma=2$. 
Lower panel: The dependence for different spectral indices for the energy bin 1-10 GeV.
All curves are measured for PSR J0007+7303.
}
\end{figure}

{\bf 2b)}
The results of our analysis for the faint halo 
around PSR J1513-5908 are displayed in Fig. \ref{fig:sim_1513_faint}.
Since we fixed the flux in the energy bin 10-100 GeV, the signal 
depends drastically upon the spectral index.
If the halo is observed at $TS\sim 100$ in the energy range 10-100 GeV,
then one may expect a signal with similar significance at 1-10 GeV
in the case of soft spectra 
(see the upper panel of Fig. \ref{fig:sim_1513_faint} for $\Gamma=2.4$). 
In the case of hard spectra ($\Gamma=2$, lower panel of Fig. \ref{fig:sim_1513_faint})
the signal is observed in both the 1-10 GeV and 100-500 GeV bins with similar significance $TS\sim 40$.

\subsection{TS-flux scaling}
\label{app:tss}

In order to put constraints on the halo luminosity we studied the dependence 
of test statistics for halos upon their fluxes (or, equivalently, the number of photons).
We sampled $\sim 10$ halo flux values for each energy bin 
and generated events for these fluxes with {\it gtobssim}. 
As above, we assumed two choices for the spectral index:
$\Gamma~=~2$ and $\Gamma~=~2.4$.
Then we processed these events with {\it gtlike} and took corresponding $TS$ values from the maximums of the obtained TS curves.

Let us first discuss the case of PSR J0007+7303. 
In the upper panel of Fig.~\ref{fig:ts_scal}
we show our results for $\Gamma=2$. Assuming the ansatz $TS=a F^b$ we obtained the following scaling:
\be
\label{eq:tscta}
\begin{split}
TS_{1-10}\simeq & 100 \left[\frac{F^{1-10\;\mathrm{GeV}}}{4.6\times 10^{-9}\;\mathrm{ph/cm}^2\mathrm{s}}\right]^{b_1}\,,\\
&~~~~~~~~~~~~~~~~~~~~~~~~~~~~\, b_1=1.54\pm 0.06\,,\\
TS_{10-100}\simeq & 100 \left[\frac{F^{10-100\;\mathrm{GeV}}}{5.7\times 10^{-10}\;\mathrm{ph/cm}^2\mathrm{s}}\right]^{b_2}\,,\\
&~~~~~~~~~~~~~~~~~~~~~~~~~~~~\, b_2=1.42\pm 0.14\,,\\
TS_{100-500}\simeq & 100 \left[\frac{F^{100-500\;\mathrm{GeV}}}{2.4\times 10^{-10}\;\mathrm{ph/cm}^2\mathrm{s}}\right]^{b_3}\,,\\
&~~~~~~~~~~~~~~~~~~~~~~~~~~~~\, b_3=1.33\pm 0.10\,. 
\end{split}
\ee

For the case of $\Gamma=2.4$ we found, essentially, the same scaling as Eq.~\eqref{eq:tscta};
see the lower panel of Fig.~\ref{fig:ts_scal}. For the 1-10 GeV bin we have
\be
 \begin{split}
TS^{\Gamma=2.4}_{1-10}\simeq & 100 \left[\frac{F^{1-10\;\mathrm{GeV}}}{5.1\times 10^{-9}\;\mathrm{ph/cm}^2\mathrm{s}}\right]^{b'}\,,\\
&~~~~~~~~~~~~~~~~~~~~~~~~~~~~\, b'=1.52\pm 0.13\,
\end{split}
\ee

We also found that in each energy bin 
this scaling depends on the background flux
(galactic interstellar and isotropic emission), but this dependence is very mild and 
can notably affect the scaling only for extreme values, which are ruled out by observations.

As for the case of PSR J1513-5908, in the energy bins 10-100 GeV and 100-500 GeV 
we found almost the same scaling as that for PSR J0007+7303, 
while the scaling in the energy bin 1-10 GeV is very different 
from that obtained in Eq.~\eqref{eq:tscta}.
For instance, in order to detect a halo in this energy bin at $TS=100$ one requires the flux 
$F^{1-10\;\mathrm{GeV}}\sim 2\times 10^{-8}\;\mathrm{ph/cm}^2\mathrm{s}$,
which is 25 times bigger than the analogous flux in the case of PSR J0007+7303.

On the other hand, the scaling at energies above 10 GeV is essentially the same for 
both pulsars, which suggests that if the angular size of a halo is larger  than the  LAT PSF,
the scaling of halo test statistics with the flux in each energy bin 
is a generic property which is valid for any source and
can be used to derive constraints from the data.

We additionally simulated a faint halo around the pulsar PSR J1709-4429 and
found, up to a few percent difference, the same $TS$ dependence on the 
flux as Eq.~\eqref{eq:tscta}.
We also performed additional checks to verify that the scaling \eqref{eq:tscta} 
is valid with accuracy $\lesssim 20\%$
in the region of interest $TS\sim 50$ for the energy bins 1-10 GeV and 10-100 GeV for various spectral indices
and background fluxes.

Overall, our analysis implies that in the energy bins 10-100 GeV and 100-500 GeV 
the scaling is given by Eq.~\eqref{eq:tscta} and is valid 
for both subpopulations. 
The scaling in the energy bin 1-10 GeV [Eq.~\eqref{eq:tscta}] is generic only
for the pulsars with large halos.

\section{Fluctuations or not?}
\label{app:fluct}

\begin{table}[h!]
\begin{tabular}{|c|c|c|c|c|c|c|c|}
\hline
 {\it N-S} source & $N_{exp}$ & $N_{obs}$ & $N_{2008-2011}$& p-val\\\hline
1 & 28.161 & 29 & 20 & 0.611\\\hline
2 & 30.606 & 27&  22& 0.294\\\hline
3 & 76.535 & 63& 55& 0.065\\\hline
4 & 24.820 & 23& 18& 0.408\\\hline
5 & 42.622 & 37& 31& 0.219\\\hline
6 & 20.560 & 25& 15& 0.861\\\hline
7 & 5.391 & 7& 4&0.823\\\hline
8 & 6.984 & 7& 6&0.601\\\hline
9 & 8.281 & 7& 7&0.414\\\hline
10 & 13.017 & 5& 11&0.011\\\hline
11 & 12.476 & 11& 10&0.408\\\hline
12 & 27.747 & 34& 21&0.897\\\hline
13 & 23.871 & 39& 18&0.998\\\hline
14 & 40.240 & 40& 30&0.527\\\hline
15 & 33.771 & 21& 25&0.013\\\hline
16 & 39.861 & 34& 29&0.200\\\hline
17 & 19.229 & 15& 14&0.200\\\hline
18 & 15.502 & 11& 11&0.154\\\hline
\end{tabular}
\caption{
For each extended source of Ref.~\cite{Neronov:2012kz} we display
$N_{exp}$ [the expected number of photons in the time span 
October 2011-July 2015
computed using Eq.~\eqref{eq:nexp}],
$N_{obs}$ (the observed number of photons),
$N_{2008-2011}$ (the number of photons observed in the span 
August 2008-October 2011), 
p-val (the Poissonian p-values corresponding to the probability to observe 
$N \le N_{obs}$ events expecting $ N_{exp}$). 
}
\label{tab:nssours}
\end{table}

In Ref.~\cite{Neronov:2012kz} the LAT events above 100 GeV 
from August 2008 to October 2011 were analyzed using the minimal spanning tree method.
Then only the halos coincident with known sources from the TeVCat catalogue
were selected for further analysis. 
This procedure, however, does not guarantee that the {\it N-S} sources selected in
that way are not due to background fluctuations.
In this section we perform an independent check to make sure that this 
is not the case.

For each source listed in Table II of Ref.~\cite{Neronov:2012kz}
we computed the expected number of photons in the time span 
October 2011-July 2015
inside the circles corresponding to the halo sizes
($\theta_{90}$ from Ref.~\cite{Neronov:2012kz}).
For each source we assumed the fluxes as retrieved from the data part 
August 2008 - October 2011, 
which was used in Ref.~\cite{Neronov:2012kz}. 
This yielded the following expected number of
photons:
\be 
\label{eq:nexp}
N_{exp}=N_{2008-2011}\frac{\mathcal{E}_{2011-2015}}{\mathcal{E}_{2008-2011}}\,,
\ee
where $N_{2008-2011}$ is the number of photons observed inside the $\theta_{90}$
circles from August 2008 to October 2011,
and by $\mathcal{E}$ we denote the exposition for the relevant time span. 
Notice that in Table II of Ref.~\cite{Neronov:2012kz} $N_{ph}$ is the background subtracted number of photons. The number $N_{2008-2011}$ we present here also includes the background ones.

Having computed the expected number of photons  we compared them
to $N_{obs}$, the observed numbers of photons in the time 
span October 2011-July 2015
inside the same halos. 
The results are shown in Table~\ref{tab:nssours}. 
For each source we computed the
p-values corresponding to the Poissonian probability to observe 
$N \le N_{obs}$ events expecting $ N_{exp}$. 
The p-values are compatible with our null hypothesis, that is, the {\it N-S} sources 
have stable fluxes and are not produced by fluctuations.

\section{3FGL sources}
\label{app:3fgl}

In this appendix we show the best-fit results for 
3FGL sources within our $10^\circ$ RoI 
for the 1-10 GeV bin for PSR J0007+7303. 
Corresponding parameters are listed in Table~\ref{tab:fit_param}. 
The best fits for other extended models can be obtained upon request 
at   \texttt{mikhail.ivanov@cern.ch}. For source model definitions 
see Ref.~\cite{fermimodels}.

\newpage 

\onecolumngrid

\begin{table}[h!]
\centering
\caption{\label{tab:fit_param} Results of the $gtlike$ fit for the model that includes a $1.1^{\circ}$ uniform halo around 
PSR J0007+7303. 
Benchmark values from the 3FGL catalogue are presented for comparison. 
The benchmark value for the normalization of galactic and isotropic emissions is 1.} 
\begin{tabular}{|c|c|c|c|c|}
\hline
 {\small 3FGL name} & Model and parameters &   Parameters, 3FGL & Parameters with halo    & Distance, $^{\circ}$\\
   &  &  &   &    \\
\hline\hline
J0007.0+7302 & PLSuperExpCutoff, ($E_c$(MeV), & 1732, & 1734$\pm$10, & 0.0 \\
&  $N_0 \times 10^{10}$, $\gamma_1$)& 1.45, $-$1.208 & 1.3464$\pm$0.0067, $-$1.1860$\pm$0.0048  &\\
J0012.4+7040 & PowerLaw, ($N_0 \times 10^{13}$, $\gamma$) & 5.5, $-$2.48 & Removed ($TS<$ 5)  & 2.41 \\
J0028.6+7507 & PowerLaw, ($N_0 \times 10^{13}$, $\gamma$) & 5.04, $-$2.34 & 5.00$\pm$0.44, $-$2.32$\pm$0.09  & 2.54 \\
J2355.4+6939 & PowerLaw, ($N_0 \times 10^{13}$, $\gamma$) & 6.62, $-$2.54 & Removed ($TS<$ 5)  & 3.52 \\
J0008.5+6853 & LogParabola ($N_0 \times 10^{12}$, $\alpha$, $\beta$) & 4.26,  2.42, 0.93 & 1.96$\pm$0.19, 2.11$\pm$0.10, 0.403$\pm$0.072  & 4.16 \\
J2356.9+6812 & PowerLaw, ($N_0 \times 10^{12}$, $\gamma$) & 1.67,  $-$2.63 & 1.14$\pm$0.17, $-$2.77$\pm$0.14  & 4.90 \\
J0004.2+6757 & PowerLaw, ($N_0 \times 10^{13}$, $\gamma$) & 6.01, $-$2.49 & 7.13$\pm$0.75, $-$4.35$\pm$0.46  & 5.09 \\
J2353.3+6639 & LogParabola ($N_0 \times 10^{12}$, $\alpha$, $\beta$) & 9.12,  2.45, 0.999 & 1.69$\pm$0.38, 2.67$\pm$0.16, 0.013$\pm$0.057  & 6.49 \\
J0116.8+6913 & PowerLaw, ($N_0 \times 10^{12}$, $\gamma$)  & 4.73,  $-$2.75  & 26.87$\pm$5.81, $-$4.81$\pm$0.23  & 6.77 \\
J0008.7+6558 & LogParabola ($N_0 \times 10^{11}$, $\alpha$, $\beta$) & 1.45,  2.50, 0.999 & 19.16$\pm$6.37, 3.26$\pm$0.49, 5.61$\pm$0.71  & 7.08 \\
J0110.2+6806 & PowerLaw, ($N_0 \times 10^{13}$, $\gamma$)  & 1.95,  $-$1.99  & 2.066$\pm$0.081, $-$1.862$\pm$0.064  & 7.17 \\
J0000.1+6545 & PowerLaw, ($N_0 \times 10^{12}$, $\gamma$) & 1.00, $-$2.41 & 1.70$\pm$0.76, $-$4.96$\pm$1.69  & 7.32 \\
J2340.7+8016 & PowerLaw, ($N_0 \times 10^{13}$, $\gamma$)  & 5.68,  $-$1.91  & 6.66$\pm$0.64, $-$3.04$\pm$0.19  & 7.37 \\
J0152.8+7517 & PowerLaw, ($N_0 \times 10^{14}$, $\gamma$)  & 1.11,  $-$1.77  & 1.04$\pm$0.13, $-$1.50$\pm$0.19  & 7.48 \\
J0135.0+6927 & PowerLaw, ($N_0 \times 10^{13}$, $\gamma$)  & 9.57,  $-$2.55  & Removed ($TS<$ 5)  & 7.86 \\
J0153.4+7114 & PowerLaw, ($N_0 \times 10^{15}$, $\gamma$) & 2.31,  $-$1.56 & 121.80$\pm$36.24, $-$1.96$\pm$0.13  & 8.28 \\
J0204.0+7234 & PowerLaw, ($N_0 \times 10^{13}$, $\gamma$) & 3.95,  $-$2.22 & 0.026$\pm$0.60, $-$0.045$\pm$0.39  &  8.56\\
J2355.5+8154 & PowerLaw, ($N_0 \times 10^{11}$, $\gamma$) & 1.0  $-$2.86 & 3.95$\pm$1.23, $-$0.63$\pm$0.13  &  8.87\\
J0025.7+6404 & PowerLaw, ($N_0 \times 10^{14}$, $\gamma$) & 4.47,  $-$2.08 & Removed ($TS<$ 5)  & 9.13 \\
J0051.6+6445 & PowerLaw, ($N_0 \times 10^{13}$, $\gamma$) & 2.56,  $-$2.28 & 4.72$\pm$121.35, $-$4.61$\pm$32.16  & 9.17 \\
J0217.5+7349 & PowerLaw, ($N_0 \times 10^{11}$, $\gamma$) & 6.13,  $-$2.90 & 5.58$\pm$138.71, $-$9.03$\pm$1.73  &  9.21\\
J0001.0+6314 & PowerLaw, ($N_0 \times 10^{12}$, $\gamma$) & 8.62,  $-$2.73 & 1100$\pm$97, $-$0.63$\pm$0.05  & 9.82 \\
galactic & Diffuse, (prefactor) & -  &  0.9842$\pm$0.0018  &  - \\
isotropic & Diffuse, (normalisation) & -  &  0.933$\pm$ 0.024  &  - \\
\hline
\hline
\end{tabular}
\end{table}

\twocolumngrid

\end{document}